\RequirePackage{fix-cm}
\documentclass[twocolumn,epjc3]{svjour3}  
\smartqed  % flush right qed marks, e.g. at end of proof
\RequirePackage{graphicx}
\usepackage{color}
\journalname{Eur. Phys. J. A}
\def\j{{(j)}}
\def\nng{n_{{\rm g}n}}
\def\nnp{n_{{\rm g}p}}

\begin{document}

\title{
Light clusters in the liquid proto-neutron star inner crust
}


\author{H. Dinh Thi\thanksref{e1,addr1}
		\and
		A. F. Fantina\thanksref{e2,addr2}
        \and
        F. Gulminelli\thanksref{e3,addr1} %etc.
}

\thankstext{e1}{e-mail: dinh@lpccaen.in2p3.fr}
\thankstext{e2}{e-mail: anthea.fantina@ganil.fr}
\thankstext{e3}{e-mail: gulminelli@lpccaen.in2p3.fr}


\institute{Université de Caen Normandie, ENSICAEN, CNRS/IN2P3, LPC Caen UMR6534, F-14000 Caen, France \label{addr1}
           \and
           Grand Acc\'el\'erateur National d’Ions Lourds (GANIL), CEA/DRF - CNRS/IN2P3, Boulevard Henri Becquerel, 14076
Caen, France \label{addr2}
}

\date{Received: date / Accepted: date}

\maketitle

\begin{abstract}
Being born hot from core-collapse supernova, the crust of the proto-neutron star is expected to be made of a Coulomb liquid and composed of an ensemble of different nuclear species.
In this work, we study the beta-equilibrated proto-neutron-star crust in the liquid phase in a self-consistent multi-component approach, employing a compressible liquid-drop description of the ions including the ion centre-of-mass motion. 
Particular care is also devoted to the calculation of the rearrangement term, thus ensuring thermodynamic consistency.
We compare the results of the multi-component plasma calculations with those obtained within a one-component (single-nucleus) approach, showing that important differences arise between the predictions of the two treatments.
In particular, the abundances of helium clusters become important using a complete multi-component plasma approach, and eventually dominate the whole distribution at higher temperature in the crust. 
%\fg{This phenomenon can be viewed as the stellar matter counterpart of the well known $\alpha$-clustering predicted in dilute symmetric nuclear matter \cite{Schuck1,Schuck2}.}

\end{abstract}
\section{Introduction}
\label{sec:intro}

The inner crust of a neutron star (NS) is a region of about 1~km thickness laying a few hundreds metres below the surface of the star \cite{hpy2007}. 
Despite being relatively thin, this layer is expected to play a crucial role in several NS phenomena, such as cooling and transport properties, see e.g. Refs.~\cite{Newton2013a,Horowitz2015,Lin2020,Schmitt2018}. 

Usually, the composition of the inner crust is determined under the so-called ``cold-catalysed matter'' hypothesis, where matter is supposed to be at zero temperature and composed of a periodic lattice of fully ionised and neutron-rich ions, surrounded by unbound neutrons, in beta equilibrium with a highly relativistic electron gas. 
However, this zero-temperature picture may not reflect the realistic composition of the NS crust \cite{hpy2007,Chamel2008,Goriely2011}. 
Indeed, NSs are born hot from gravitational core-collapse supernova, with initial temperature exceeding $10^{10}$~K \cite{hpy2007}. 
In these conditions, the proto-NS (PNS) crust is expected to be made of a Coulomb liquid, characterised by a statistical distributions of different nuclear species immersed in a background of neutrons and protons and neutralised by an electron gas (see e.g Ref.~\cite{Oertel2017} for a review). 
As the star cools down, the temperature decreases and the width of the nuclear distribution becomes narrower (see e.g. Refs.~\cite{Fantina2020,Carreau2020b} for a recent investigation), until eventually the zero-temperature ground state is reached\footnote{Actually, depending on cooling rates, the crust could be frozen at a temperature higher than the crystallisation one (see e.g. Ref.~\cite{Goriely2011}), so that the ground-state composition may in fact not be attained.}.
Therefore, at sufficiently low temperatures, where the distribution is very peaked, the mixture of different nuclei can be approximated by the most probable one, i.e. the most favourable from the thermodynamic point of view.
Even if this one-component plasma (OCP) picture can be justified in such regimes, the presence of impurities induced by the coexistence of different nuclear species in the multi-component plasma (MCP) can have a strong impact on the PNS crust transport and evolution properties (see e.g. Refs.~\cite{Pons2013,Vigano2013}) and needs to be taken into account.
For this reason, NS (cooling) simulations usually recast the presence of the nuclear distribution in the so-called impurity parameter, defined as the variance of the cluster proton number distribution and often ad-hoc adjusted on cooling data.
Consistent calculations of this parameter at the crystallisation temperature for the NS crust were carried out in Refs.~\cite{Fantina2020,Carreau2020b} and were very recently extended to higher density and temperature regimes in the inner crust in Ref.~\cite{Dinh2022_mcp}. 

More generally, the presence of different clusters and systematic comparisons between the OCP and MCP predictions were previously investigated in several works.
For example, in Refs.~\cite{Burrows1984,Souza2009,Hempel2010,Blinnikov2011} conditions relevant for core-collapse supernova were explored, while in Ref.~\cite{gulrad2015} several stellar conditions encountered in both supernova and NS matter were studied.
The common outcome of these works was that the thermodynamic properties of matter, such as pressure and entropy, can be described relative well with the OCP approach. 
On the contrary, as for the composition, the ion mass numbers are systematically overestimated by the OCP, while MCP calculations also predict a considerable amount of light clusters, that even tend to prevail over heavy ones as the density and temperature increase. 

To overcome the limitation of OCP approaches to account for the presence of light degrees of freedom, several works, mainly devoted to supernova matter, added explicitly light clusters on top of the representative heavy cluster.
For example, alpha particles were included in Refs.~\cite{lattimer1991,Shen1998}, $^3$H and $^3$He were added in Ref.~\cite{OConnor2007}, while deuterons, tritons, and helions were considered in Refs.~\cite{Avancini2012,Avancini2017,Pais2015,Custodio2021}.
However, in beta equilibrium conditions, more neutron-rich light isotopes are expected to appear \cite{gulrad2015}, thus in principle a full distribution should be taken into account.

In our previous investigations \cite{Dinh2022}, we studied the PNS crust in the liquid phase in the OCP approach.
To this aim, we modelled the crust using a compressible liquid drop (CLD) model as in Refs.~\cite{Carreau2019,Carreau2020,dinh2021,dinhEPJA21}, with parameters optimised on microscopic energy-density functionals and surface properties fitted from the experimental nuclear masses. 
Even though this method is not as microscopic as a full density functional treatment, it was recently shown to provide results in good agreement with extended Thomas-Fermi calculations both at zero \cite{Grams2022} and finite temperature \cite{Carreau2020}.
At the same time, it also allows to perform calculations at a reduced numerical cost with respect to more microscopic approaches, such as self-consistent Hartree-Fock (HF), Hartree-Fock+BCS or Hartree-Fock-Bogoliubov (HFB), or extended Thomas-Fermi plus Strutinski Integral methods (e.g., \cite{MagHen2002,Grill2011,Baldo2007,Gogelein2007,Shelley2020,Pearson2018,Pearson2020,Pearson2022,Shelley2021}).
Our findings showed that the inclusion of the ion centre-of-mass motion is important for the description of the PNS composition; namely, the appearance of heavy or light clusters depends on the competition between the finite-size and the translational contribution to the free energy.
Specifically, the latter term tends to favour the appearance of lighter, essentially unbound clusters near the crust-core transition.
In our subsequent upcoming work \cite{Dinh2022_mcp}, we studied the liquid crust in the MCP approach, and show notable differences in the prediction of the PNS composition with respect to that obtained with a OCP approach, particularly in what concerns the abundance of light nuclei.

In this work, we pursue the study of the PNS inner crust in beta equilibrium, as achieved in late (P)NS cooling stages, in the temperature regime between 1~MeV and 2~MeV, which is believed to be high enough for the crust to be in the liquid phase \cite{hpy2007,Carreau2020}.
We present here additional comparisons of the results obtained using the OCP approximation and a fully self-consistent MCP approach.

The paper is organised as follows.
Section~\ref{sec:OCP} is devoted to present the OCP approximation: in Sect.~\ref{sec:OCP-model} the CLD formalism within the OCP approximation is described; in particular, the different contribution to the ion energetics are detailed in Sect.~\ref{sec:fcluster}, while the system of variational minimisation equations are discussed in Sect.~\ref{sec:crust_OCP}; numerical results are discussed in Sect.~\ref{sec:OCP-results}.
In Sect.~\ref{sec:MCP}, we introduce the self-consistent MCP approach: after recalling the derivation for the nuclear distribution in Sect.~\ref{sec:distribution}, we discuss the numerical results and the comparison with the OCP findings in Sect.~\ref{sec:MCP-results}.
Finally, conclusions are given in Sect.~\ref{sec:conlusions}.

%--------------------------------------------------------------------
\section{One-component plasma approximation}
\label{sec:OCP}

\subsection{Model of the crust in the OCP approach}
\label{sec:OCP-model}

The modelling of the NS inner crust in the liquid phase within the OCP approximation was described in details in Ref.~\cite{Dinh2022}. 
In the following, we recall the main points and assumptions. 

In the OCP approximation, the NS inner crust at each given thermodynamics condition is assumed to be composed of identical Wigner-Seitz (WS) cells, each of which has a volume of $V_{\rm WS}$ and contains a fully ionised ion of mass number $A$, proton number $Z$, and volume $V_N$, immersed in a uniform background of electrons, neutrons, and protons whose densities are denoted by $n_e$, $\nng$, and $\nnp$, respectively. 
The total baryonic density $n_{\rm g}$ and isospin asymmetry $\delta_{\rm g}$ of the nuclear medium are thus $n_{\rm g} = n_{{\rm g}n} + n_{{\rm g}p}$, and $\delta_{\rm g} =\frac{n_{{\rm g} n} - n_{{\rm g}p}}{n_{{\rm g}n} + n_{{\rm g}p}} $.

In the liquid phase, the ion is not localised at the centre of the cell, as in the solid phase, but it can explore the whole volume. 
Nevertheless, because of the OCP approximation, a WS volume can still be defined as the optimal volume encompassing each (moving) ion, and is obtained from the condition of charge neutrality (see Eq.~(\ref{eq:charge})). 

At a total baryonic density $n_B$ and temperature $T$, the total free-energy density of the system is given by:
    \begin{equation}
        \mathcal{F} = \mathcal{F}_e +  \mathcal{F}_{\rm g} (1-u) + \frac{F_i}{V_{\rm WS}}, 
         \label{eq:total_free_energy_density}
    \end{equation}
where $\mathcal{F}_e =\mathcal{F}_e(n_e, T)$ denotes the electron free-energy density\footnote{We use the uppercase letter $F$, lowercase letter $f$, and notation $\mathcal{F}$ to denote the free-energy per ion, free-energy per nucleon, and free-energy density, respectively.}, for which we use Eq.~(2.65) in Ref.~\cite{hpy2007}, and $\mathcal{F}_{\rm g} =\mathcal{F}_B(n_{\rm g}, \delta_{\rm g}, T) + m_n c^2 n_{{\rm g}n} + m_p c^2 n_{{\rm g}p}$ is the free-energy density (including the nucleon rest masses) of the surrounding neutron and proton gases, with $m_n$ and $m_p$ being the neutron and proton mass, respectively, and $\mathcal{F}_B(n, \delta, T)$ being the nuclear matter free-energy density at baryonic density $n$, isospin $\delta$, and temperature $T$ (see also the discussion in the next section).
The term $-u\mathcal{F}_{\rm g}$, with $u=\frac{V_N}{V_{\rm WS}}$ being the cluster to the WS cell volume ratio, is an excluded-volume term accounting for the subtraction of the gas states from the nuclear partition sum, hence avoiding double counting of unbound single-particle states ~\cite{gulrad2015,Mallik2021}. 
Finally, $F_{i}$ is the cluster free energy, which is described in the following section.

\subsubsection{Cluster free energy in the liquid phase} \label{sec:fcluster}
In this work, we calculate the cluster free energy within a CLD model approach, as in \cite{Carreau2019,Carreau2020,dinh2021,dinhEPJA21}. Therefore, $F_i$ in Eq.~(\ref{eq:total_free_energy_density}) can be decomposed into:
    \begin{equation}
          F_i = M_i c^2 + F_{\rm bulk} + F_{\rm Coul + surf +  curv} + F^{\star}_{\rm trans},
          \label{eq:Fi0}
    \end{equation}
where $M_i = (A-Z)m_n + Zm_p$ denotes the cluster bare mass, $F_{\rm bulk} = \frac{A}{n_i} \mathcal{F}_B(n_i, I, T)$ represents the bulk free energy of the cluster with internal density $n_i = \frac{A}{V_N}$ and isospin  asymmetry $I = \frac{A-2Z}{A}$. 
To obtain the free-energy density of homogeneous nuclear matter $\mathcal{F}_B$,  we employed the self-consistent mean-field thermodynamics \cite{lattimer1985,Ducoin2007} (see also Sect.~3.1 in Ref.~\cite{Dinh2022} for more details). 
Within this approach, the free-energy density of a nuclear matter system of total density $n$ and isospin asymmetry $\delta$ at temperature $T$, $\mathcal{F}_B(n,\delta,T)$, is written as the sum of a ``potential'' and a ``kinetic'' part as follows:
    \begin{equation}
        \mathcal{F}_B(n,\delta,T) = \mathcal{F}_{\rm kin}(n,\delta,T) + \mathcal{V}_{\rm MM}(n,\delta) \ .
     \label{eq:Fb}
    \end{equation}

The kinetic term, $\mathcal{F}_{\rm kin}$, carries the temperature dependence and can be written as:
    \begin{equation}
        \mathcal{F}_{\rm kin} = \sum_{q=n,p} \left[ \frac{-2 k_{\rm B} T}{\lambda_q^{3/2}} F_{3/2}\left( \frac{\tilde{\mu}_q}{k_{\rm B} T} \right) + n_q \tilde{\mu}_q  \right] \ ,
    \label{eq:Fkin}
    \end{equation}
where $q=n,p$ denotes neutrons and protons, $k_{\rm B}$ is the Boltzmann constant, and $F_{3/2}$ is the Fermi-Dirac integral. 
It is to be noted that the nucleon thermal wavelength, $\lambda_q = \left( \frac{2 \pi \hbar^2}{k_{\rm B} T m^\star_q} \right)^{1/2}$, is related to the density-dependent neutron and proton effective mass, $m^\star_q = m^\star_q(n, \delta)$, due to the non-locality of the effective nucleon-nucleon interaction. 
Furthermore, the term $\tilde{\mu}_q$ in Eq.~(\ref{eq:Fkin}) is an auxiliary chemical potential, related to the thermodynamical one, $\mu_q$, by: 
\begin{equation}
    \tilde{\mu}_q = \mu_q - U_q.
\end{equation}
where $U_q$ is the mean-field potential.

To describe the potential term, we used the so-called meta-modelling approach, as proposed in Refs.~\cite{Margueron2018a,Margueron2018b}, where $\mathcal{V}_{\rm MM}$ can be written as a Taylor expansion truncated at order $N$  around the saturation point ($n=n_{\rm sat}$, $\delta=0$):
    \begin{equation}
        \mathcal{V}^{N}_{\rm MM}(n, \delta) = \sum_{k=0}^{N} \frac{n}{k!}(v^{\rm is}_{k} +v^{\rm iv}_{k}\delta^2 )x^{k}u^{N = 4}_{k}(x),
         \label{eq:vMM}
    \end{equation}
where $x = (n - n_{\rm sat})/(3n_{\rm sat})$, $u^{N}_{k}(x)$ is a function ensuring the convergence at zero-density limit, and the parameters $v^{\rm is}_{k}$ and $v^{\rm iv}_{k}$ can be expressed as linear combinations of the nuclear matter empirical parameters (see Ref.~\cite{Margueron2018a} for details). 
It was shown that realistic functionals at zero temperature can be reproduced with high precision truncating the expansion at the order $N = 4$. 
To illustrate the results, in this work, we use the empirical parameters corresponding to the BSk24 \cite{BSK24} functional, that was shown to be consistent with current data from astrophysical and nuclear physics constraints \cite{dinh2021}.  

The finite-size contribution, namely the Coulomb, surface and curvature energies, $F_{\rm Coul+surf+curv}$ in Eq.~(\ref{eq:Fi0}), is given by:
    \begin{equation}
        F_{\rm Coul + surf +  curv} = V_{\rm WS}(\mathcal{F}_{\rm Coul} + \mathcal{F}_{\rm surf} + \mathcal{F}_{\rm curv})\ .
    \end{equation}
In the WS cell approximation, the Coulomb energy density can be expressed as:
    \begin{equation}
        \mathcal{F}_{\rm Coul}  = 2\pi (en_ir_N)^2\left(\frac{1-I}{2} -\frac{n_{{\rm g}p}}{n_i}\right)^2 u \eta_{\rm Coul}(u), \label{eq:Fcoul}
    \end{equation}
where $e$ is the elementary charge, $r_N$ is the cluster radius, and  $\eta_{\rm Coul}(u)$ is a decreasing function of $u$ accounting for the electron screening. 
For spherical nuclei, $\eta_{\rm Coul}$ %has the following expression 
reads \cite{Ravenhall1983_prl,Pethick1995}:
    \begin{equation}
        \eta_{\rm Coul}(u) =\frac{1}{5}\left [ u+ 2     \left ( 1- \frac{3}{2}u^{1/3} \right ) \right ]. \label{eq:eta_Coul}
    \end{equation}
As for the surface and curvature terms, we used the same expression as in Refs.~\cite{lattimer1991}, \cite{Maru2005}, and \cite{Newton2013}:
    \begin{equation}
        \mathcal{F}_{\rm {surf}} + \mathcal{F}_{\rm {curv}} =\frac{3u}{r_N}\left ( \sigma_{\rm s}(I, T) +\frac{2\sigma_{\rm c}(I, T)}{r_N}\right ) , \label{eq:interface}   
    \end{equation}
in which $\sigma_{\rm s}$ and $\sigma_{\rm c}$ are the surface and curvature tensions. 
For $\sigma_{\rm s} (I, T = 0)$ and $\sigma_{\rm c}(I, T=0)$, we employed the parameterization proposed in Ref.~\cite{Ravenhall1983}, based on the Thomas-Fermi calculations at extreme isospin asymmetries:
    \begin{eqnarray}
        \sigma_{\rm s}(I, T = 0) &=& \sigma_0\frac{2^{p+1}+b_s}{y_p^{-p}+b_s+(1-y_p)^{-p}} \ , \label{eq:surface} \\
        \sigma_{\rm c}(I, T = 0)&=&5.5 \, \sigma_{\rm s}(I, T = 0) \frac{\sigma_{0, {\rm c}}}{\sigma_0}(\beta-y_p)\ , \label{eq:curvature} \ 
    \end{eqnarray}
where $y_p = (1-I)/2$ is the cluster proton fraction, and $(\sigma_0,\sigma_{0,{\rm c}},b_s,\beta,p)$ are the surface and curvature parameters. The parameter $p$, which is relevant only at extreme isospin value, is fixed to the canonical value $p=3$ \cite{Carreau2019}, while the others are optimised consistently with the bulk properties to reproduce the experimental nuclear masses in the Atomic Mass Evaluation (AME) 2016 table \cite{AME2016} (see Table~2 in Ref.~\cite{dinh2021}). 

In the finite-temperature regime, the surface tension is modified with respect to the case of zero temperature as \cite{lattimer1991}:
    \begin{equation}
        \sigma_{\rm s,c}(I, T) = \sigma_{\rm s,c}(I, T =0)h(T),
                \label{eq:sigmas_T}
    \end{equation}
where
\begin{equation}
h(T) = \left \{
    \begin{array}{r c l}
     0 &     &\mbox{if $T > T_c$} \ ,  \\
        \left[1 -\left(\frac{T}{T_c}\right)^2\right]^2 &     &\mbox{if $T \leq T_c$} \ ,
        \end{array}
        \right .
    \end{equation}
and $T_c$ is the critical temperature given by Eq.~(2.31) in Ref.~\cite{lattimer1991}.

Finally, the last term on the right-hand-side of Eq.~(\ref{eq:Fi0}), $F^{\star}_{\rm trans}$, is the translational free energy, which accounts for the cluster centre-of-mass motion. % when the crust is in the liquid phase. 
Although this term is usually neglected \cite{Avancini2017,Avancini2009,Shen2011}, or excluded from the minimisation with respect to the cluster size \cite{lattimer1991,Schneider}, as discussed in Ref.~\cite{Dinh2022}, it should be taken into account in calculating the crust composition at temperatures above the crystallisation point.
Indeed, we showed \cite{Dinh2022} that the translational energy can have a considerable impact on the equilibrium configuration of the crust, especially if the beta-equilibrium condition holds, as it is assumed in the present work (see also Sect. \ref{sec:crust_OCP}). 

%For a system of  non-relativistic classical particles, 
For non-relativistic ions and accounting for in-medium effects (see Sect.~3.3 in Ref.~\cite{Dinh2022} for details), the translational free energy, $F^{\star}_{\rm trans}$, can be written as:
    \begin{equation}
        {F}_{\rm trans}^\star= k_{\rm B} T\ln \left( \frac{1}{V_{\rm f}}\frac{\lambda_i^{\star 3}}{g_s}\right)  - k_{\rm B} T, 
    \label{eq:ftrans_eff}
    \end{equation}
where $g_s$ is the spin degeneracy, which is set to unity, $V_{\rm f}$ is the ``free'' volume available for the ion motion, for which we adopted the expression proposed in Ref. \cite{Lattimer1985},
    \begin{equation}
         V_{\rm f} = V_{\rm f}^{\rm OCP} = \frac{4}{3} \pi (r_{\rm WS} - r_{N})^3 \ ,
    \end{equation}
with $r_{\rm WS}$ being the WS cell radius, and $\lambda^{\star}_i = \sqrt{\frac{2\pi \hbar^2}{M^{\star}_i k_{\rm B} T}}$ is the ion thermal wavelength, with $M_i^\star$ being the ion effective mass.
If the effects from the cluster size and from the nucleon background are neglected, then $V_{\rm f} = V_{\rm WS}$, $M_i^\star = M_i$, and the expression of the translational free energy reduced to that of ideal gas, as given in Eq.~(2.71) in Ref.~\cite{hpy2007}.
However, in-medium effects are expected not to be negligible in the NS inner crust, especially at high densities, near the crust-core transition.
Indeed, as the density increases, the cluster volume fraction $u$ tends to increase, thus restricting the volume available for the cluster motion, and the nucleon background becomes denser, thus its effect cannot be neglected.

The influence of the surrounding nuclear medium can be recasted into an ion effective mass, that has been obtained by solving the hydrodynamic equation assuming that the nuclear matter inside and outside of the cluster is an incompressible, irrotational, and non-viscous fluid, as proposed in Refs.~\cite{Epstein1988,Sedrakian1996,Magierski2004,Magierski2004b,Martin2016,Chamel2017} for a different application.
This leads to the following expression for the ion effective mass (see Sect.~2 in Ref.~\cite{Dinh2022} for details):
    \begin{equation}
         M_i^{\star} = M_i \left[ 1 - \delta^{\rm f} + \frac{(\delta^{\rm f}-\gamma)^2}{\delta^{\rm f} + 2\gamma} \right] \ ,
    \label{eq:m*-sphere}
    \end{equation}
where $\gamma = \frac{\nng m_n + \nnp m_p}{M_i/V_N}$ is the mass-density ratio of the surrounding medium to the cluster, and $\delta^{\rm f}$ is the mass fraction of nucleons inside the ion participating in the superfluid flow. 
%If the ion is moving in vacuum, then $\gamma = 0$, yielding the correct limit, $M_i^{\star} = M_i$, as one expects. 

Assuming $m_n \approx m_p$ and that the density of free protons is negligible at $k_{\rm B}T \leq 2$~MeV \cite{Dinh2022} (see also Fig.~\ref{fig:Ypgas}) leads to $\gamma \approx \nng / n_i$. 
To precisely evaluate the ion effective mass, fully microscopic calculations of the dynamical properties of nuclei in a dense nuclear-matter medium would be needed.
Nevertheless, one can estimate the effect of the effective mass in some limiting cases:
%Regarding $\delta^{\rm f}$, we consider the following cases:
%
    \begin{enumerate}
        \item  First, we consider $\delta^{\rm f}=0$. In this picture, the ion is completely impermeable to the surrounding nuclear medium:
        \begin{equation}
            M_i^\star = M_i \left( 1 + \frac{1}{2} \gamma \right) \ ,
             \label{eq:m*-hard-sphere}
        \end{equation}
        and the expression for the effective mass of an impenetrable hard sphere given in Ref.~\cite{Sedrakian1996} is recovered.
        \item Secondly, we consider that the ion is partially permeable and that all the neutrons occupying continuum states participate to the flow. Therefore, $\delta^{\rm f} = \gamma \approx \frac{\nng}{n_i}$, and the effective mass reads:
        \begin{equation}
        M^\star \approx  M_i \left(1- \frac{\nng}{n_i}\right) \ ,
         \label{eq:m*-deltamin}
        \end{equation}
        which is equal to the mass in the e-cluster representation of Ref.~\cite{gulrad2015}. 
        \item In the third case, we consider $\delta^{\rm f} = 1 - Z/A = N/A$, meaning that all neutrons in the cluster participate in the (external) flow. The effective mass can be thus written as:
        \begin{equation}
            M^\star \approx M_i \left[ y_p + \frac{(1-y_p-\nng/n_i)^2}{1-y_p + 2 \nng/n_i} \right] \ .
        \label{eq:m*-deltamax}
        \end{equation}
        \item Finally, we consider the extreme case $\delta^{\rm f} = 1$, meaning that the ion is fully permeable to the flow of the outside nucleon fluid. % thus all neutrons in the cluster participate in the ``free'' motion. 
        This picture corresponds to the calculation in Ref.~\cite{Magierski2004b}, leading to:
        \begin{equation}
             M_i^\star = M \frac{(1-\gamma)^2}{1+2\gamma} \ .
        \label{eq:m*-mb}
        \end{equation}
    \end{enumerate}
The influence of these prescriptions of the ion effective mass on the crust equilibrium composition is illustrated in Sect.~\ref{sec:OCP-results}.

\subsubsection{Crust composition in the OCP approximation} \label{sec:crust_OCP}
To obtain the optimal configuration of the crust at each given density $n_B$ and temperature $T$, the total free-energy density  
$\mathcal{F}$ in Eq.~(\ref{eq:total_free_energy_density}) 
is minimised under the constraints of baryon number conservation and charge neutrality:
 \begin{eqnarray}
     n_B &=& \frac{2n_p}{(1-I)}\left(1 - \frac{n_{{\rm g}n} + n_{{\rm g}p}}{n_i}\right) + n_{{\rm g}n} + n_{{\rm g}p},\label{eq:baryon_conservation}\\
     n_e &=& n_p + n_{{\rm g}p}\left(1- \frac{2n_p}{n_i (1-I)}\right),
        \label{eq:charge}
 \end{eqnarray}
where $n_p = Z/V_{\rm WS}$. One can show that the function to be minimised can be written as \cite{Dinh2022}:
  \begin{eqnarray}
            \Omega&=& \frac{2n_p}{1-I}\frac{F_i}{A}+ \left[ 1 - \frac{2n_p}{(1-I)n_i}\right]\mathcal{F}_{\rm g} + \mathcal{F}_e  \nonumber \\
             &+& \gamma_1 \left[n_B - \frac{2n_p}{1-I}\left(1 - \frac{n_{{\rm g}n} + n_{{\rm g}p}}{n_i}\right) - n_{{\rm g}n} -n_{{\rm g}p}\right] \nonumber \\
             &+& \gamma_2 \left[n_e - n_p - n_{{\rm g}p}\left(1- \frac{2n_p}{n_i (1-I)}\right)\right],
        \label{eq:Omega}
\end{eqnarray}
where $\gamma_1$ and $\gamma_2$ are the two Lagrange multipliers, which can be shown to be directly connected to the neutron and proton chemical potentials: 
    \begin{eqnarray}
        \gamma_1 &=&  \frac{\partial \mathcal{F}_{\rm g}}{\partial n_{{\rm g}n}}  + \frac{2n_pn_i}{n_i(1-I)- 2n_p} \frac{\partial(F_i/A)}{\partial n_{{\rm g}n}} \equiv \mu_n, \label{eq:mu_n}   \\
        \gamma_1 + \gamma_2 &= & \frac{\partial \mathcal{F}_{\rm g}}{\partial n_{{\rm g}p}}  + \frac{2n_pn_i}{n_i(1-I)- 2n_p} \frac{\partial(F_i/A)}{\partial n_{{\rm g}p}} \equiv \mu_p. \label{eq:mu_p}
    \end{eqnarray}
Identifying $\mu_{{\rm g}n ({\rm g}p)} \equiv \left.\frac{\partial \mathcal{F_{\rm g}}}{\partial n_{{\rm g}n ({\rm g}p)}}\right|_{n_{{\rm g}p ({\rm g}n) }}$ as the chemical potentials of the unbound neutrons (protons), the second term on the right-hand-side of Eqs.~(\ref{eq:mu_n})-(\ref{eq:mu_p}) accounts for the in-medium modification induced by the centre-of-mass translation and the Coulomb interaction.

At beta equilibrium, we minimised the function $\Omega$ defined in Eq.~(\ref{eq:Omega}) with respect to the variational variables ($r_N$, $I$, $n_i$, $n_p$, $n_e$), and obtained the following system of coupled equations:
\begin{eqnarray}
    \mu_{p} + \mu_e &=& \mu_{n},    \label{eqocp:betaequi} \nonumber \\
     \frac{\partial (F_i/A)}{\partial r_N} &=& 0, \label{eqocp:rn}\\
    n_i^2\frac{\partial}{\partial n_i}\left(\frac{F_i}{A}\right) &=& P_{\rm g} , \label{eqocp:ni}\\
    \frac{F_i}{A} + (1-I) \frac{\partial}{\partial I}\left(\frac{F_i}{A}\right) &=& \mu_n - \frac{P_{\rm g}}{n_i}, \label{eqocp:I}\\
     2\left[  \frac{\partial}{\partial I}\left(\frac{F_i}{A}\right) - \frac{n_p}{1-I}  \frac{\partial}{\partial n_p}\left(\frac{F_i}{A}\right) \right] &=& \mu_n - \mu_p, \label{eqocp:Iandnp}
\end{eqnarray}
where $\mu_e = \partial \mathcal{F}_e/\partial n_e$ is the electron chemical potential, and $P_{\rm g} = \mu_{n} n_{{\rm g}n} +\mu_{p} n_{{\rm g}p} - \mathcal{F}_{\rm g} $ is the pressure of the dripped nucleons (including in-medium effects, see Eqs.~(\ref{eq:mu_n})-(\ref{eq:mu_p})).
At each given thermodynamic condition, this system of five coupled differential equations, Eqs.~(\ref{eqocp:betaequi})-(\ref{eqocp:Iandnp}), is solved numerically together with Eqs.~(\ref{eq:baryon_conservation}) and (\ref{eq:charge}), yielding the equilibrium crust composition. % in beta equilibrium.

\subsection{Numerical results: impact of the ion centre-of-mass motion}
\label{sec:OCP-results}

\begin{figure}[htb!]
        \centering
        \includegraphics[scale = 0.45]{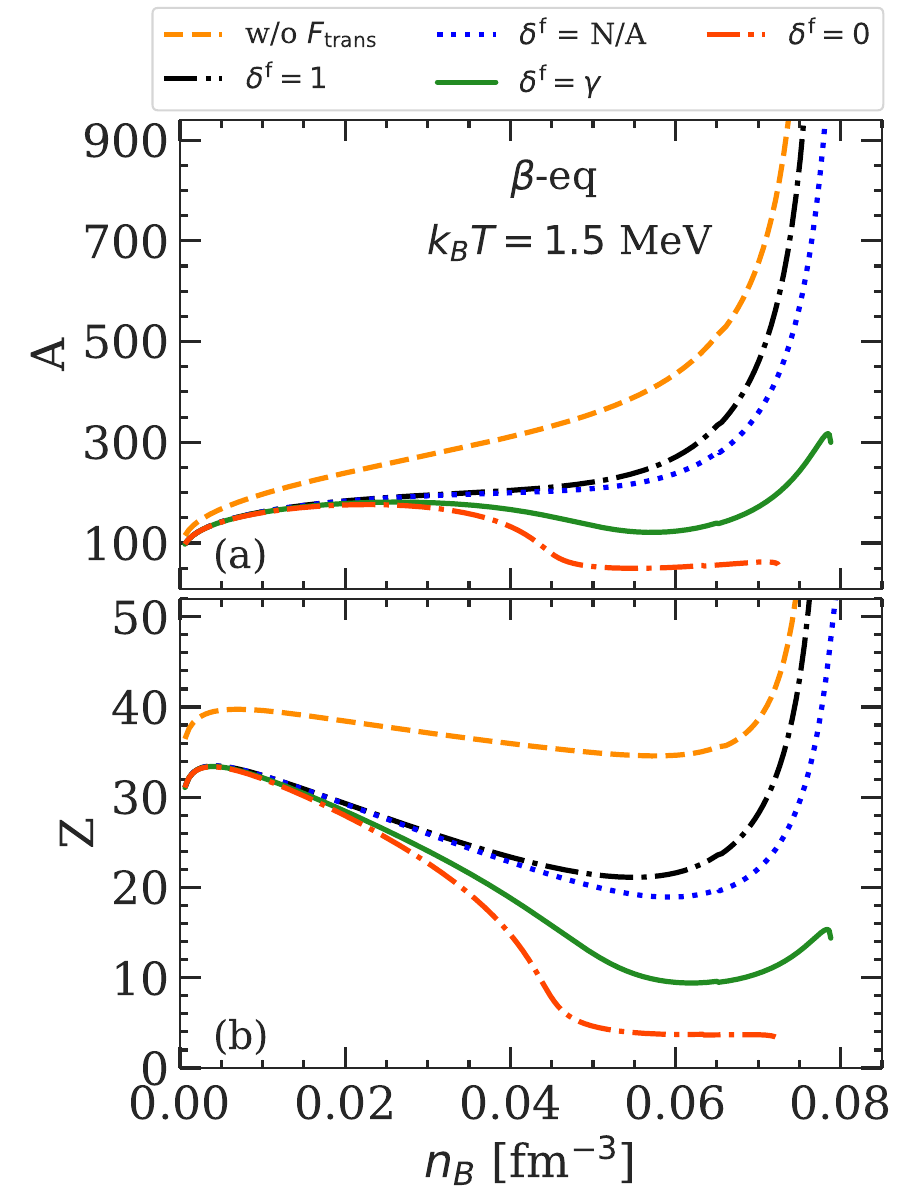}
        \caption{Cluster mass number $A$ (panel a) and proton number $Z$ (panel b) as a function of the total baryonic number density $n_B$ at $k_{\rm B} T = 1.5$~MeV in the OCP approximation, using different prescriptions for the ion effective mass. See text for details.}
        \label{fig:betaqui_AZvsnb_bsk24_4meff_1p5MeV}
    \end{figure}

We start by discussing the influence of different prescriptions of the ion effective mass, that enters the translational free energy via Eq.~(\ref{eq:ftrans_eff}), on the composition.
In Fig. \ref{fig:betaqui_AZvsnb_bsk24_4meff_1p5MeV}, we display the mass number $A$ (panel a) and proton number $Z$ (panel b) as a function of the total baryonic density $n_B$ at a chosen temperature, $k_{\rm B}T = 1.5$~MeV. 
The dashed orange curve is obtained without the translational term, while the other results are obtained including $F^{\star}_{\rm trans}$ with different prescriptions for $M_i^\star$, corresponding to the values of $\delta^{\rm f}$ listed in Sect.~\ref{sec:OCP-model}: $\delta^{\rm f} = 1$ (dash-dotted black lines), $\delta^{\rm f} = N/A$ (dotted blue lines), $\delta^{\rm f} = \gamma$ (solid green lines), and $\delta^{\rm f} = 0$ (dash-dotted red lines). 
In general, the translational free energy reduces both the mass and proton numbers of the cluster, and  the effect is more significant with decreasing $\delta^{\rm f}$. 
In particular, without the translational term, the proton number $Z$ remains almost constant, $Z \approx 40$, for almost all densities in the inner crust. 
On the other hand, when the cluster centre-of-mass motion is considered, the proton number is reduced by more than ten units. In the most extreme case, $Z \approx 4$ at $n_B \geq 0.05$ fm$^{-3}$ if the ion is considered as a hard sphere ($\delta^{\rm f} = 0$, dash-dotted red line). This can be understood as follows. 
As discussed in Ref.~\cite{Dinh2022}, in the variational procedure, the cluster size is determined from the minimisation of $\Omega$ with respect to $r_N$, that is, Eq.~(\ref{eqocp:rn}). 
Substituting Eq.~(\ref{eq:Fi0}) in Eq.~(\ref{eqocp:rn}), we can easily show that:
\begin{equation}
        \frac{\partial (f_{\rm Coul + surf +  curv} + f_{\rm trans}^\star)}{\partial r_N} = 0.
        \label{eq:rn1}
\end{equation}  
Without the translational term, Eq.~(\ref{eq:rn1}) reduces to the well-known Baym virial theorem \cite{BBP1971} with an additional curvature term, $\mathcal{F}_{\rm surf} + 2\mathcal{F}_{\rm curv} = 2\mathcal{F}_{\rm Coul}$. 
We can observe from Fig.~\ref{fig:betaqui_AZvsnb_bsk24_4meff_1p5MeV} that this equation results in large cluster, $A > 100$ (dashed orange line in panel a). 
On the other hand, if we consider only $f^{\star}_{\rm trans}$, then we can show that this function has a minimum at considerable smaller $A$ than that obtained from the virial theorem. As a result, when $f_{\rm trans}^{\star}$ is included, then the optimal values of $A$ and $Z$ are shifted to lower values.

\begin{figure}[htb!]
    \centering
    \includegraphics[scale = 0.45]{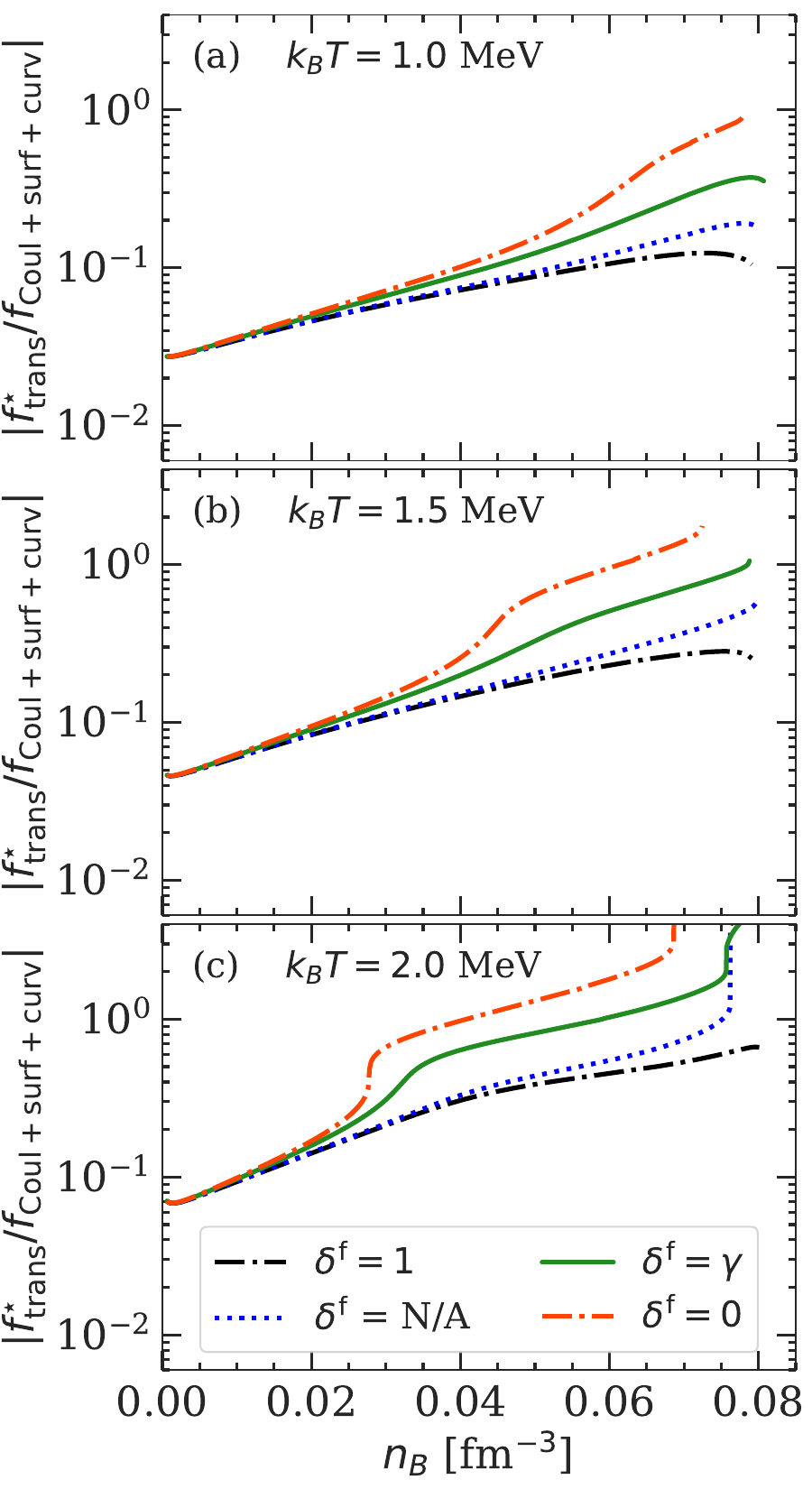}
    \caption{Absolute value of the ratio of the translational free energy per nucleon to the sum of surface, curvature, and Coulomb free energies per nucleon as a function of the total baryonic number density $n_B$ at three different temperatures: $k_{\rm B} T = 1.0$ MeV (panel a), $k_{\rm B} T = 1.5$ MeV (panel b), and $k_{\rm B} T = 2.0$ MeV (panel c) in the OCP approximation. Different prescriptions for the ion effective mass, corresponding to different values of $\delta^{\rm}$, are considered. See text for details.}
    \label{fig:ratio_4meff}
\end{figure}

\begin{figure}[htb!]
    \centering
    \includegraphics[scale =  0.45]{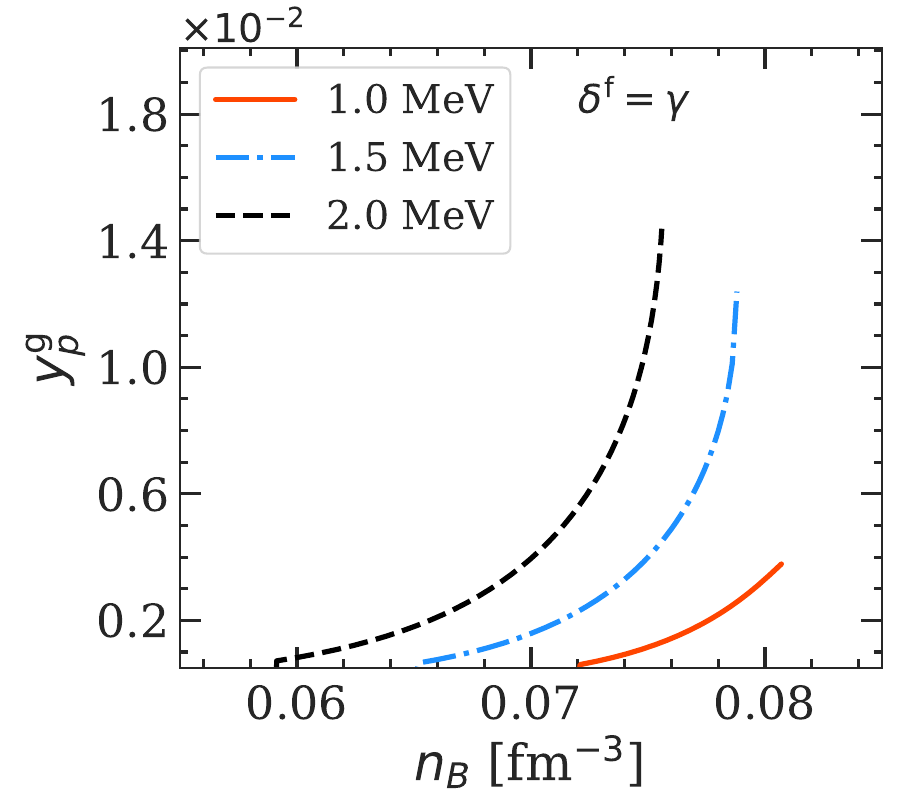}
    \caption{Proton fraction in the gas, $y_p^{\rm g} = (1 - \delta_{\rm g})/2$ at three different temperatures: $k_{\rm B}T = 1.0$~MeV (solid red line), $k_{\rm B}T = 1.5$~MeV (dash-dotted blue line), and $k_{\rm B}T = 2.0$~MeV (dashed black line), obtained in the OCP approximation using $F_{\rm trans}^{\star}$ with $\delta^{\rm f} = \gamma$.}
    \label{fig:Ypgas}
\end{figure}

Moreover, from Fig.~\ref{fig:betaqui_AZvsnb_bsk24_4meff_1p5MeV}, we can also see that the influence of the translational degrees of freedom becomes more prominent as the density increases. 
This behaviour can be understood by looking at Fig.~\ref{fig:ratio_4meff}, that displays the absolute ratio $\left|\frac{f_{\rm trans}^{\star}}{f_{\rm Coul+surf+curv}}\right|$ as a function of the total baryonic density $n_B$ for three different temperatures: $k_{\rm B}T = 1.0$~MeV (panel a), $k_{\rm B}T = 1.5$~MeV (panel b), and $k_{\rm B}T = 2.0$~MeV (panel c), and for different prescriptions of the ion effective mass. 
One can observe that the ratio increases with densities for all the temperatures and $\delta^{\rm f}$ considered. 
This is because in beta equilibrium matter becomes more neutron rich as the density increases. 
Therefore, the contribution from the Coulomb, surface, and curvature term decreases and the impact from $f_{\rm trans}^{\star}$ becomes stronger. 
Moreover, one can see that the contribution from the translational free energy of the ion increases with increasing effective mass, or equivalently decreasing $\delta^{\rm f}$, and this behaviour is also reflected in the $A$ and $Z$ trend exhibited in Fig.~\ref{fig:betaqui_AZvsnb_bsk24_4meff_1p5MeV}. 
However, the discrepancy in the results due to different effective mass prescriptions manifest itself at $n_B > 0.02$~fm$^{-3}$. 
At lower densities, both the free energy ratio (Fig.~\ref{fig:ratio_4meff}) and the equilibrium composition (Fig.~\ref{fig:betaqui_AZvsnb_bsk24_4meff_1p5MeV}) are very similar in the four cases. 
This is because in the outer layer of the inner crust, the nucleon gas densities are relative low. 
Therefore, the ion mass is not significantly modified by the medium, and $M_i^{\star} \approx M_i$. %, $ \forall \delta^{\rm f}$. 
Furthermore, in this region, the ions are very far from each other, and therefore $V_{\rm f} \approx V_{\rm WS}$. 
Therefore, the results obtained with different expressions of $f_{\rm trans}^{\star}$ are closer to that of the ideal-gas expression (see Fig.~7 of Ref.~\cite{Dinh2022}). 
Additionally, the ratio $\left|\frac{f_{\rm trans}^{\star}}{f_{\rm Coul+surf+curv}}\right|$, thus the effect from the centre-of-mass motion, increases with temperature. 
As a result, one can expect that clusters become smaller at lower densities as temperature increases, as shown in Ref.~\cite{Dinh2022}. 
This means that even at the OCP level, at sufficiently high densities and temperatures, the light degrees of freedom can become more energetically favorable than the heavier ones and dominate the crust composition until they are totally dissolved into homogeneous nuclear matter (see e.g. Fig.~7 in Ref.~\cite{Dinh2022}).

In the prescription of the ion effective mass described in Sect.~\ref{sec:OCP-model}, we have assumed that the proton gas be negligible.
We now show that this assumption remains valid throughout the PNS crust at the temperature of interest for this study.
In Fig.~\ref{fig:Ypgas}, we plot the fraction of free protons in the nucleon background, $y_p^{\rm g} = (1 - \delta _{\rm g})/2$, as a function of $n_B$ for three selected temperatures, and for $\delta^{\rm f} = \gamma$, that is, neutrons in the cluster occupying the continuum states are those participating to the (external) flow. 
Indeed, we believe that this picture is the most realistic one among those that we have considered here.
We can observe that the free protons start to emerge at high density in the inner crust, close to the crust-core transition. 
Moreover, the higher the temperature, the sooner the proton gas appears, as expected. 
Specifically, the protons start to drip at $n_B \approx 0.072$ fm$^{-3}$, $n_B \approx 0.065$ fm$^{-3}$, and $n_B \approx 0.060$ fm$^{-3}$, for $k_{\rm B}T = 1.0$~MeV, $k_{\rm B}T = 1.5$~MeV, and $k_{\rm B}T = 2.0$~MeV, respectively. 
Although $y_p^{\rm g}$ increases with temperature, at the highest temperature that we consider, $k_{\rm B}T = 2.0$~MeV, the fraction of free proton is still less than 1.4$\%$. 
Therefore, the presence of the proton gas can be safely neglected, which also justify ``a posteriori'' our choice of approximating $\gamma \approx \nng/n_i$. %and the approximation we used for $\gamma$ is a good approximation. 

As already mentioned in the introduction, the OCP approximation can be adopted at relatively low temperatures and densities, however, this assumption becomes less reliable at higher temperatures or as far as the presence of impurities are concerned.
For this reason, we calculated the PNS crust in the MCP approach, which will be discussed in the next section.

\section{Multi-component plasma approach}
\label{sec:MCP}

\subsection{Nuclear distribution in the MCP approach}
\label{sec:distribution}

The detailed derivation of the MCP formalism is presented in our upcoming work \cite{Dinh2022_mcp}; here, we only recall the main points.
In the MCP approach, at each thermodynamic condition, the NS crust is composed of different WS cells containing different nuclear species $\j \equiv (A^\j, Z^\j)$, each one having a frequency of occurrence $p_j$, such that $\sum_j p_j = 1$. 
The neutron and electron gases are assumed to be homogeneous in all cells, and therefore, $\nng^\j = \nng$, and $n_e^\j = n_e$. 
In principle, at finite temperature the proton gas should also be considered, but as shown in Fig.~\ref{fig:Ypgas} the proton gas fraction remains negligible in almost all densities in the PNS crust.
Moreover, the inclusion of the proton gas introduces an additional equation, which adds cost to the numerical computation in a fully self-consistent MCP calculation.
Therefore, in this section, we neglect the presence of the proton gas, i.e. we set $n_{{\rm g}p} = 0$. 
In addition, charge neutrality condition is supposed to hold in each cell,
\begin{equation}
    n_e = n_p = n_p^{(j)} = \frac{Z^{(j)}}{V_{\rm WS}^{(j)}}.
    \label{eqMCP:ne_j}
\end{equation}
As a result, we can neglect the Coulomb correlation among different cells, and the free energy of a cell $\j$ does not depend explicitly on other components\footnote{We note that there is an implicit dependence of the cell free energy on the other cells through charge neutrality, see Eq.~(\ref{eq:rearr}) and related discussion.} $\j' \neq \j$.

Let $n_N^\j =  p_j / \langle V_{\rm WS}\rangle$, with $\langle \rangle$ indicating the ensemble averages, be the ion density of a species $\j$; the total free energy density of the system thus reads:
    \begin{equation}
        \mathcal{F^{\rm MCP}} = \sum_j n_N^{(j)} \left( F_i^\j -V_N^\j\mathcal{F}_{\rm g} \right) + \mathcal{F}_{\rm g} + \mathcal{F}_e  \ ,
     \label{eqMCP:FMCP}
    \end{equation}
where the neutron gas free-energy density is
\begin{equation}
\mathcal{F}_{{\rm g}} =\mathcal{F}_B(\nng, \delta_{\rm g} = 1 , T) + m_n c^2 \nng \ .
\end{equation}

The cluster free energy $F_i^\j$ has the same form as in Eq.~(\ref{eq:Fi0}), except for the translational free energy.
Indeed, in the MCP picture, the centre-of-mass position of the different ions is not confined to the single WS cell but clusters can explore the whole average volume, this correlation between the different ion species introducing a breaking of the linear-mixing rule \cite{Fantina2020,Dinh2022_mcp,gulrad2015}. %``free'' volume, $ \langle V_{\rm f} \rangle$. 
As in the expression for the translational energy in the OCP approximation, Eq.~(\ref{eq:ftrans_eff}), two in-medium effects have to be considered: (i) the ions cannot actually explore the whole volume, consistently with the excluded-volume approach, and (ii) the ion mass is replaced by an effective mass.
As a result, the translation free energy for each ion reads:
    \begin{equation}
        F^{\star,\j, {\rm MCP}}_{\rm trans}  = k_{\rm B} T \left[ \ln \left( \frac{p_j}{\langle V_{\rm f}\rangle}\frac{(\lambda_i^{\star,\j}) ^3}{g_s^\j}\right)  - 1 \right] ,
    \label{eq:ftrans_MCP}
    \end{equation}
where the spin degeneracy $g_s^\j$ is set to unity for all nuclear species, and the ``free'' volume available for the ion motion, $ \langle V_{\rm f} \rangle$, is given by:
    \begin{equation}
        \langle V_{\rm f}\rangle = \langle V_{\rm WS}\rangle - \langle V_{N}\rangle \ .
    \end{equation}
This latter expression was also adopted in the statistical study by Ref.~\cite{Hempel2010} on supernova matter\footnote{In Ref.~\cite{Hempel2010}, the authors also considered the volume of free nucleons in the excluded-volume term, but this latter correction is neglected in our calculation.}.
As for the effective mass appearing in the ion thermal wavelength, we adopt for the MCP calculations the expression in Eq.~(\ref{eq:m*-deltamin}), that we consider to be the most realistic prescription among the ones listed in Sect.~\ref{sec:OCP}.

In order to calculate the nuclear distribution in the MCP approach, thus the ion densities $n_N^\j$, the thermodynamic potential in the canonical ensemble has to be minimised \cite{Fantina2020,Carreau2020b,gulrad2015,Grams2018}.
Since the electron and neutron gas free-energy densities appearing in Eq.~(\ref{eqMCP:FMCP}) are independent of $n_N^\j$, the variation with respect to the ion density can be performed on the ion part only.
However, the variations $d n_N^\j$ are not independent, because of the baryon number conservation and charge neutrality relations:
    \begin{eqnarray}
        n_B &=&  n_{{\rm g}n} + \sum_j n_N^{(j)} \left(A^{(j)} - n_{{\rm g}n} V_N^{(j)} \right) \ , \label{eqMCP:baryon}\\
	   n_e &=& n_p = \sum_j n_N^{(j)}Z^{(j)} \ . \label{eqMCP:charge}
    \end{eqnarray}
To include these two constraints in the minimisation of the total free-energy density, $\mathcal{F}^{\rm MCP}$, two Lagrange multipliers, that can be directly related to the proton and neutron chemical potentials, $\mu_p$ and 
\begin{equation}
   \mu_{n}  = \frac{\partial \mathcal{F}_{\rm g}}{\partial \nng}  
    + \frac{\sum_j n_N^\j \left(\partial F^\j_i/ \partial \nng \right)}{1-\sum_j n_N^\j V_N^\j} \ ,
\label{eq:mun_MCP}
\end{equation}
can be introduced. 
We note that the latter expression, Eq.~(\ref{eq:mun_MCP}), has the same form as the corresponding one in the OCP, Eq.~(\ref{eq:mu_n}), where we can identify the first term on the right-hand-side of the equation, $\frac{\partial \mathcal{F}_{\rm g}}{\partial \nng}$, as the chemical potential of the neutron gas $\mu_{{\rm g}n}$, while the second term accounts for the in-medium effect, that is, the dependence of the ion free energy on the neutron gas density. 
Furthermore, we can also observe that in the limiting case where the nuclear distribution contains only one species, i.e. $n_N^\j = \frac{1}{V_{\rm WS}}$, one can recover the OCP expression, Eq.~(\ref{eq:mu_n}).

It can be shown (see Ref.~\cite{Dinh2022_mcp} for details) that the resulting equation for the equilibrium densities $n_N^\j$ can be recasted in terms of a single-ion canonical
potential, defined as
    \begin{equation}
         \label{eq:omegaj}
         \Omega_i^\j = F_i^{\star,\j} - V_N^\j \mathcal{F}_{\rm g} 
        +  \mathcal{R}^\j - k_{\rm B} T \ln \bar{u}_{\rm f} \ ,
    \end{equation}
%\begin{equation}
%    \Omega_i^\j + k_{\rm B}T \ln n_N^\j + \mu_e Z^\j - \mu_{n} \left(A^{(j)} - \nng V_N^{(j)} \right) = 0 \ ,
%\end{equation}
where 
\begin{equation} 
\bar{u}_{\rm f} = \frac{\langle V_{\rm f} \rangle}{\langle V_{\rm WS} \rangle}
\label{eq:uf}
\end{equation}
is the free-volume fraction and 
\begin{equation}
        F_i^{\star,\j} = F_i^\j - F^{\star,\j, {\rm MCP}}_{\rm trans} 
        + k_{\rm B} T \ln \left(\frac{(\lambda_i^{\star, \j})^3}{g_s^\j}\right) \ ,  \label{eqMCP:Fij0star}
    \end{equation}
and from which the $n_N^\j$ can be obtained as
    \begin{equation}
        n_N^\j = \exp{\left(-\frac{\tilde{\Omega}^\j_i}{k_BT}\right)} \ ,
     \label{eqMCP:nN}
    \end{equation}
with
\begin{equation}
        \tilde{\Omega}^\j_i = \Omega_i^\j + \mu_{e} Z^\j - \mu_{n} \left(A^{(j)} - \nng V_N^{(j)}  \right) \ ,
    \label{eqMCP:tidle_omegaij}
    \end{equation}
$\mu_e = d \mathcal{F}_e/ d n_e$ being the electron chemical potential (we recall that beta equilibrium holds, thus $\mu_p = \mu_n - \mu_e$).

The $\mathcal{R}^\j$ term appearing in Eq.~(\ref{eq:omegaj}) is the so-called rearrangement term, %, which will be addressed in Sect. \ref{Sec:rearrangement}, 
arising from the dependence of the cluster free energy on the electron gas density via the Coulomb screening term.
This term is crucial in ensuring the thermodynamic consistency of the model and it has been shown that, in perturbative approaches where the chemical potentials used for the MCP calculations are approximated by those obtained within the OCP calculation, this term is necessary to recover the ensemble equivalence between the MCP and OCP approaches (see e.g. \cite{Fantina2020,Carreau2020b,Grams2018,Barros2020,Pelicer2021}).
The rearrangement term can be written as:
\begin{eqnarray}
\mathcal{R}^\j &=& Z^{(j)} \sum_{j'} n_N^{(j')}
         \frac{\partial F_{\rm Coul}^{(j')}}{\partial n_p} \nonumber \\
         &=& V_{\rm WS}^\j \bar{P}_{\rm int} \ ,
    \label{eq:rearr}
\end{eqnarray}
where 
\begin{equation}
    \bar{P}_{\rm int} = \sum_{j'} n_N^{(j')} V_{\rm WS}^{(j')}P_{\rm int}^{(j')} \ ,
    \label{eq:Pbar_int}
\end{equation}
and the pressure contributed by the Coulomb interaction is
\begin{equation}
    P_{\rm int}^{(j^\prime)} = \frac{n_p^2}{Z^{(j^\prime)}} \frac{\partial F^{(j^\prime)}_{\rm Coul}}{\partial n_p} \ .
	\label{eq:P_int}
\end{equation}

We observe that, in order to obtain $n_N^\j$, here a self-consistency problem arises, because of the implicit dependence on $n_N^\j$ of the quantities on the right-hand-side of Eq.~(\ref{eqMCP:tidle_omegaij}).
Therefore, to find the nuclear distributions in the MCP approach we simultaneously solved Eqs.~(\ref{eqMCP:baryon}), (\ref{eqMCP:charge}), (\ref{eq:mun_MCP}), (\ref{eq:uf}), and (\ref{eq:Pbar_int}).

\subsection{Numerical results: OCP vs MCP}
\label{sec:MCP-results}

\subsubsection{Crust composition}

\begin{figure*}
    \centering
    \includegraphics[scale = 0.45]{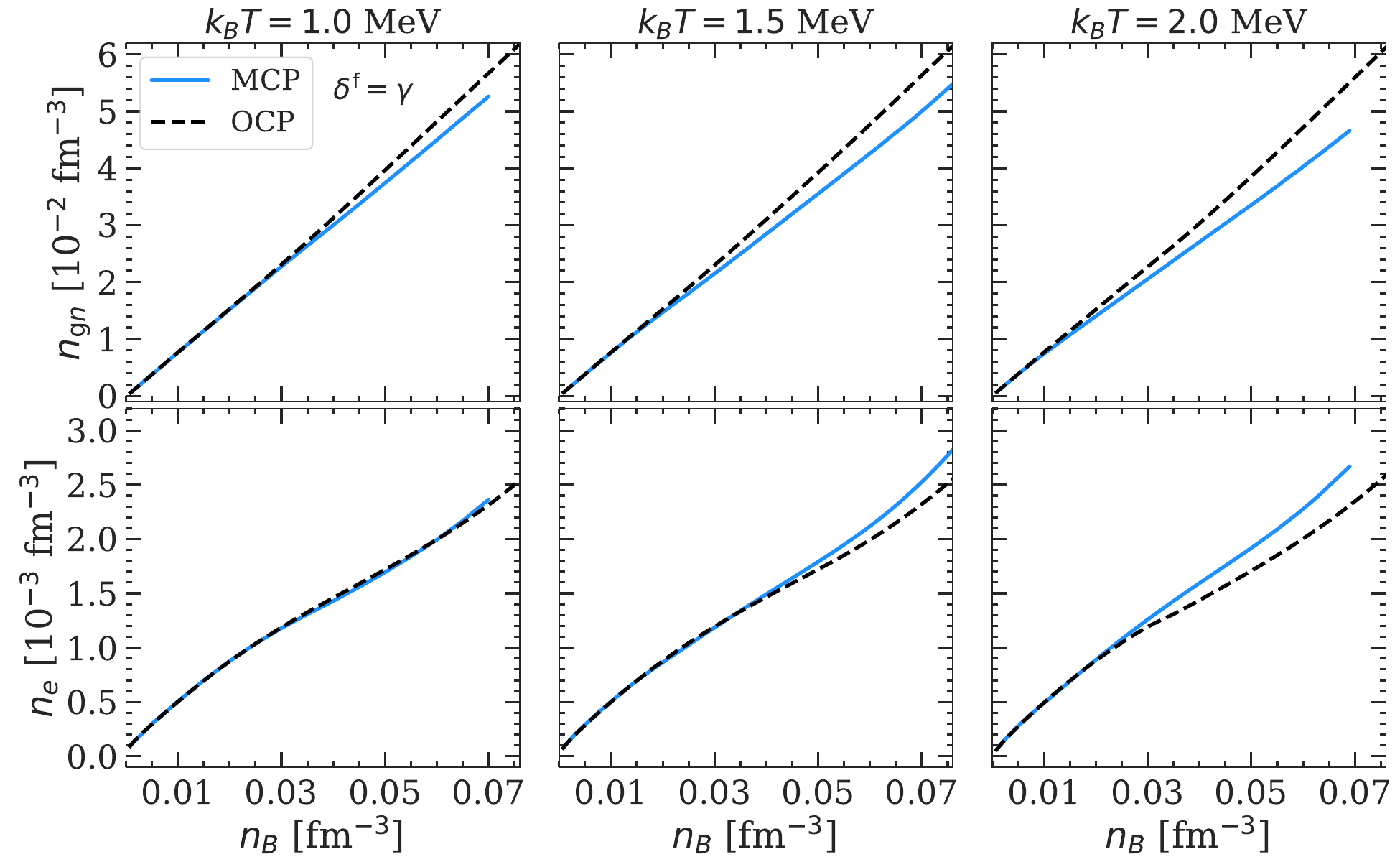}
    \caption{Neutron (upper panels) and electron (lower panels) gas densities as a function of the total baryonic density obtained in MCP (solid blue lines) and OCP (dashed black lines) calculations at three different temperatures: $k_{\rm B}T = 1.0$~MeV (left panels),  $k_{\rm B}T = 1.5$~MeV (middle panels), and  $k_{\rm B}T = 2.0$~MeV (right panels). The translational free energy is included in both cases, with $\delta^{\rm f} = \gamma$.}
    \label{fig:MCP_gas_densities}
\end{figure*}
%

%We start the discussion by comparing the prediction of the OCP and MCP results. % in what concerns the background densities.
To carry out the MCP calculations, in principle, not only all possible $(A^\j, Z^\j)$ have to be considered, but in addition in the CLD picture the ion radius $r_N^\j$ (or equivalently the ion internal density $n_i^\j$) can also fluctuate independently of the mass and proton numbers $(A^\j, Z^\j)$\footnote{We vary the ion proton number in the range $[2,100]$. For each value of  $Z$, the mass number $A$ varies in  $[2Z, 30Z]$. }. 
For each $(A, Z)$, the distribution of the ion radius can be expressed as: %,  in principle, should be taken into account:
\begin{equation} 
    p_{{A}{Z}}(r_N)=\mathcal{N}n_N(A^\j= A, Z^\j= Z,r_N) \ ,
\end{equation}
with $\mathcal{N}$ being a normalization.
However, we showed in Ref.~\cite{Dinh2022_mcp} that the distribution is typically narrow for a  given $(A,Z)$ for each thermodynamic condition $(n_B,T)$.
Therefore, in order to reduce the numerical cost, we approximated the distribution $p_{{A}{Z}}(r_N)$ by the most probable $r_N$ configuration, which can be shown to correspond to the solution of the following equation: 
    \begin{equation} \label{eq:Peq}
    P^{\star}_{\rm cl } \equiv \left.\frac{n_i^{2}}{A} \frac{\partial F_i^{\star} }{\partial n_i} \right|_{A,Z} = P_{\rm g},
    \end{equation} 
that is, the pressure equilibrium between the ion and the surrounding neutron gas.

We start by comparing the results of the OCP and MCP calculations in what concerns the background densities.
In Fig.~\ref{fig:MCP_gas_densities}, we present the neutron gas density $\nng$ (upper panels) and electron density $n_e$ (lower panels) in the MCP as a function of the total baryonic density $n_B$ at different temperatures (solid blue lines). 
For comparison, the corresponding OCP solutions are also plotted (dashed black lines). 
At all the three considered temperatures, values of $\nng$ and $n_e$ obtained with the MCP and OCP approach overlap each other in the low-density region. 
As the density and temperature increase, the discrepancy in the two approaches becomes larger.
As already noticed in the literature \cite{Fantina2020,Carreau2020b,gulrad2015}, this can be understood from the fact that when the nuclear distribution is narrow and symmetric and if the non-linear mixing term induced by the translational motion is negligible, the average quantities in MCP are close to those calculated within the OCP. 
Therefore, at low $(n_B, T)$, the gas solutions in these two approaches coincide, and the ensemble equivalence is established. 
On the other hand, at higher densities and temperatures, the distribution is spread over a wider range of $A$ and $Z$, and multiple peaks with comparable probabilities may even emerge \cite{Dinh2022_mcp,Souza2009,Hempel2010,gulrad2015,Botvina2010}, thus the symmetric shape of the distribution is no longer guaranteed. 
Furthermore, the contribution of the translational term breaking the linear-mixing rule becomes more important at high $(n_B, T)$ (see Sect.~\ref{sec:OCP}), which induces a break of the ensemble equivalence (see Ref.~\cite{Dinh2022_mcp} for details).

From Fig.~\ref{fig:MCP_gas_densities} (upper panels), we can observe that that the density of unbound neutrons in the MCP is always lower than that in the OCP approach. 
This reduction was also observed in other statistical studies, e.g. in Ref.~\cite{Burrows1984}
(see their Table 3, although this study was performed for core-collapse supernovae matter at a fixed proton fraction) and Ref.~\cite{gulrad2015} (see their Fig.~13). 
Specifically, in the latter work, that was carried out in beta equilibrium for PNS crusts, it was shown that the depletion of free nucleons is due to the appearance of light clusters (defined by $A <20$ in their calculation).
Indeed, in beta equilibrium, it is more favorable for matter to form extremely neutron-rich helium and lithium isotopes, suppressing the number of neutrons in the gas.
From these considerations, we may thus infer that the results in Fig.~\ref{fig:MCP_gas_densities} imply that at high densities and temperatures the abundance of light nuclei becomes important.

\begin{figure*}
    \centering
    \includegraphics[scale = 0.48]{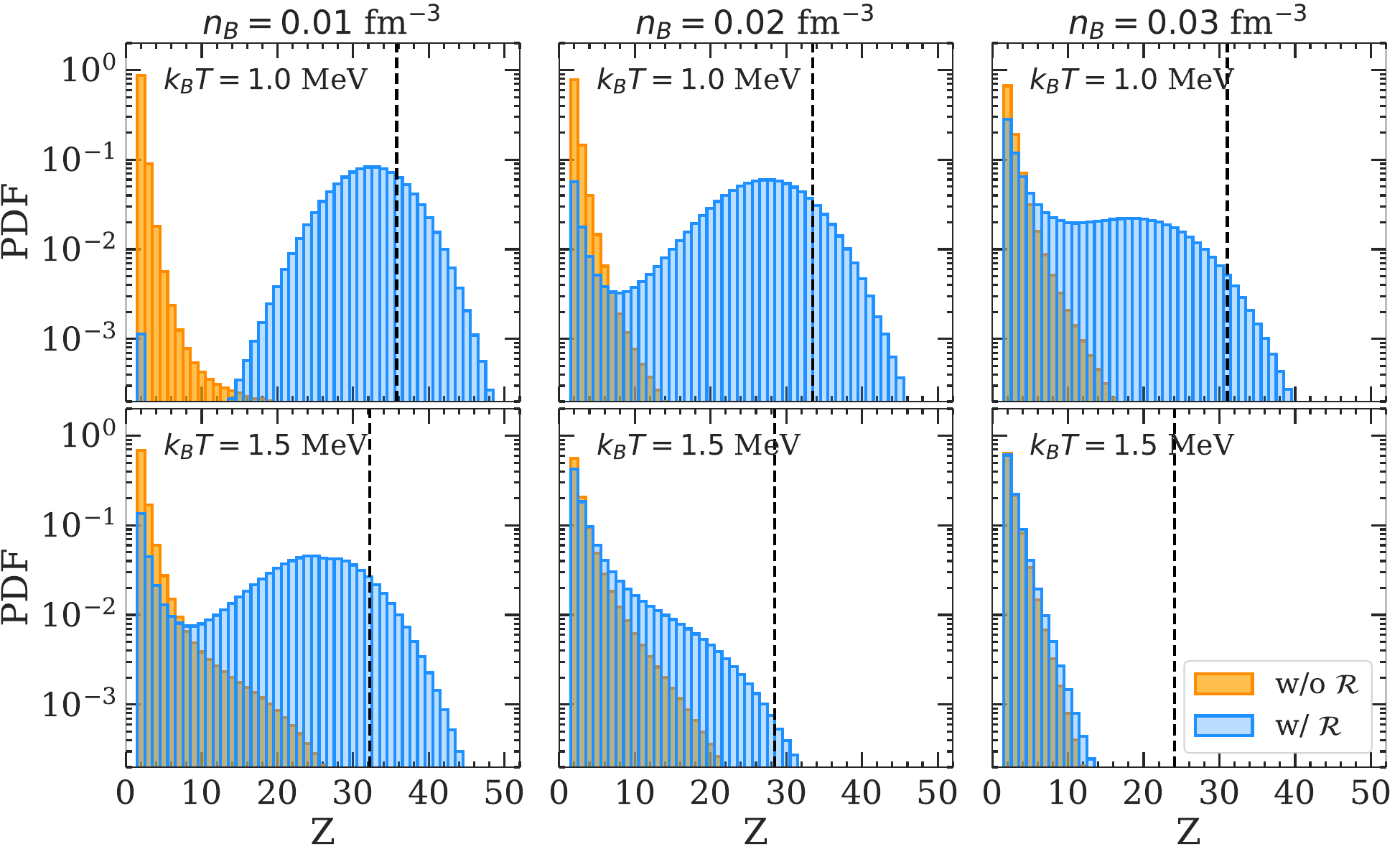}
    \caption{Probability density distribution of the ion proton number $Z$ obtained with (blue) or without (orange) the rearrangement term. Three different densities are considered: $n_B = 0.01$~fm$^{-3}$ (left panels), $n_B = 0.02$~fm$^{-3}$ (middle panels), and $n_B = 0.03$~fm$^{-3}$ (right panels). At each density, the distributions are obtained at two chosen temperatures: $k_{\rm B}T = 1.0$~MeV (upper panel) and $k_{\rm B}T = 1.5$~MeV (lower panel). The vertical dashed black line in each panel indicates the OCP solution.}
    \label{fig:MCP_PZ}
\end{figure*}

\begin{figure*}
    \centering
    \vspace*{0.8cm}
    \includegraphics[scale = 0.48]{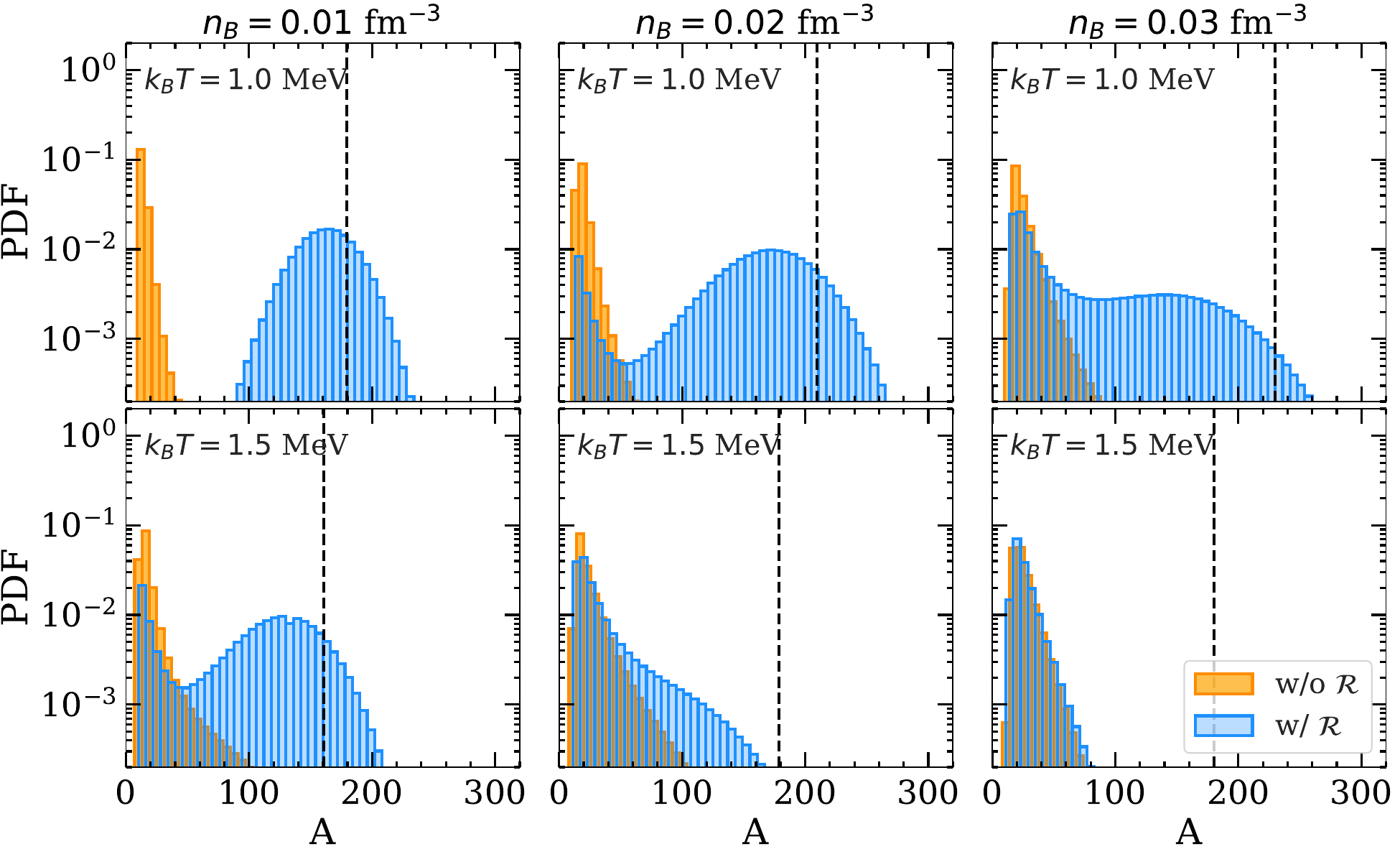}
    \caption{Same as Fig.~\ref{fig:MCP_PZ} but for the ion mass number $A$. See text for details.}
    \label{fig:MCP_PA}
\end{figure*}

To investigate further this finding, we show in Figs.~\ref{fig:MCP_PZ} and \ref{fig:MCP_PA} the probability density distributions of the proton number $Z$ and $A$, respectively at different thermodynamics conditions (blue histograms)\footnote{The probability distribution in Figs.~\ref{fig:MCP_PZ} is obtained by integrating $p_j = p_{AZ}$ over the mass number, that is, $\sum_A p_{AZ}$, while in Fig.~\ref{fig:MCP_PA} the distribution is obtained by integrating $p_{AZ}$ over the proton number, that is, $\sum_Z p_{AZ}$.}. 
The results are obtained at two chosen temperatures: $k_{\rm B}T = 1.0$~MeV (upper panels) and $k_{\rm B}T = 1.5$~MeV (lower panels), for three selected densities in the NS inner crust: $n_B = 0.01$~fm$^{-3}$ (left panels), $n_B = 0.02$~fm$^{-3}$ (middle panels), and $n_B = 0.03$~fm$^{-3}$ (right panels). 
The vertical dashed black line in each panel indicates the OCP solution. 
At the lowest density and temperature considered, that is, $n_B = 0.01$~fm$^{-3}$ and $k_{\rm B}T = 1.0$~MeV, the distributions have a Gaussian-like shape and are populated mainly by heavier clusters, with $Z > 10$ and $A>100$ (top left panels in Fig.~\ref{fig:MCP_PZ}). 
Although the most probable clusters are close to the OCP prediction (vertical lines), a slight shift towards lower $Z$ is observed for the MCP distribution. 
Interestingly, already at this rather low density, we can observe a small contribution from $Z=2$ clusters with a probability of about two orders of magnitude smaller than the most probable peak. 
As the densities increases, the peak of light clusters becomes more important (middle panels) and even prevails (right panels). 
This double-peaked distribution observed in Fig.~\ref{fig:MCP_PZ}, spanning from $Z =2$ to $Z \sim 50$, implies a large variance in $Z$, hence large impurities. 
As one may expect, the emergence of light clusters is even more noticeable at higher temperatures (see lower panels of Fig.~\ref{fig:MCP_PZ}). 
Indeed, already at the lowest density, $n_B = 0.01$~fm$^{-3}$, we can already observe a double-peak structure, and the dominance of the $Z = 2$ peak. 
As the density increases, the peak corresponding to heavier nuclei shrinks and eventually vanishes, as shown in the bottom right panel of Fig.~\ref{fig:MCP_PZ}. 
In this case, the distribution is very narrow, spanning from $Z = 2$ up to $Z \sim 10$. 
Therefore, we can expect the impurity parameter to be small (see Ref.~\cite{Dinh2022_mcp} for a more detailed discussion). 
A similar behaviour can also be observed in the distribution of the cluster mass number $A$, displayed in Fig.~\ref{fig:MCP_PA}. 
We can see that the mass number of light clusters is peaked at $A \approx 20$, thus the most probable configurations are those of extremely neutron-rich helium isotopes, as already pointed out in Ref.~\cite{gulrad2015}.
This prediction for the cluster mass number should however be taken with care, because the simple CLD approach employed here cannot account for the microscopic quartetting structure of the He wave functions and the associated Pauli blocking effects \cite{Schuck3}. 
The cluster mass number, as defined in the present work, simply accounts for the average number of neutrons that are spatially close to the clustered structure. 
The helium clusters being immersed in a neutron fluid constituted of unbound (resonant and  continuum) states, this mass number reflects the polarisation of the neutron fluid that we would expect to be induced by an $\alpha$-cluster in a microscopic calculation \cite{Schuck3}. 
This phenomenon may be viewed as the stellar matter counterpart of the well known $\alpha$-clustering predicted in dilute symmetric nuclear matter \cite{Schuck1,Schuck2}.

Comparing with the OCP prediction, in general, the latter tends to overestimate the numbers of protons and nucleons of the optimal nucleus compared to the MCP approach.
In particular, the deviation between the most probable cluster in the MCP and the one predicted by the OCP gets larger with temperature and density. 
More importantly, the OCP fails in reproducing the presence of light clusters that are instead automatically accounted for in a full MCP approach.

\begin{figure}
    \centering
    \includegraphics[scale = 0.45]{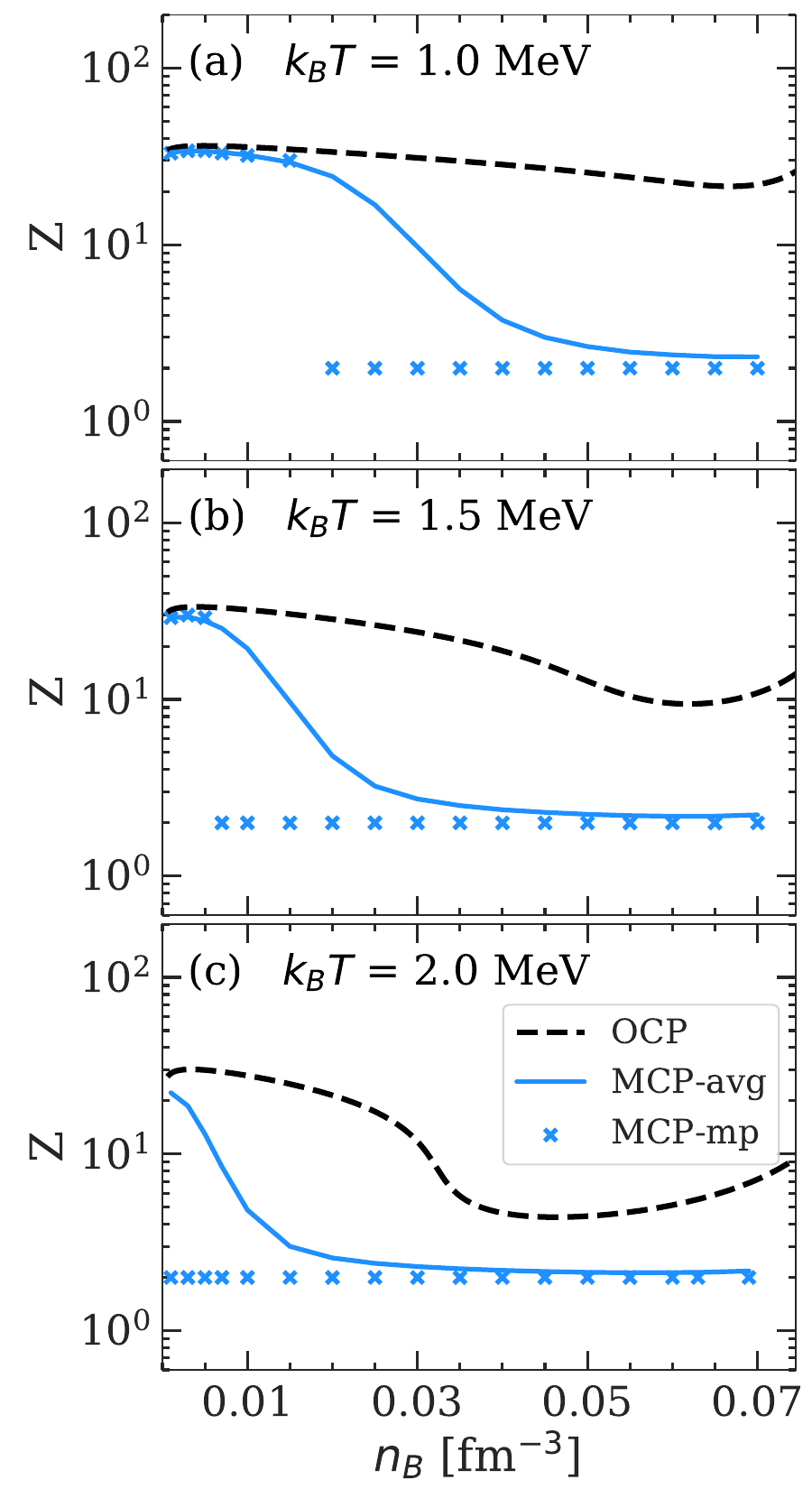}
    \caption{Average (full lines, labelled ``MCP-avg'') and most probable (crosses, labelled ``MCP-mp'') element as a function of the baryon number density $n_B$ at three different temperatures: $k_{\rm B}T=1$~MeV (panel a), $k_{\rm B}T=1.5$~MeV (panel b), and $k_{\rm B}T=2$~MeV (panel c), in the full MCP calculation. In each panel, the OCP predictions are reported as dashed black lines.}
    \label{fig:summary}
\end{figure}

The average and most probable matter composition for the full MCP and for the OCP approximation are reported in Fig.~\ref{fig:summary}.
We can see that in the full statistical calculation the He clusters are the dominant species in a large range of densities and temperatures. 
In particular, for temperatures overcoming $k_BT> 1.5$ MeV, the liquid crust can be schematically viewed as an He fluid. 
It is interesting to remark that the He dominance persists at very high density, above the predicted Mott transition density in asymmetric matter \cite{Schuck4}. 
This might be due to our simplistic approach for the cluster binding. 
However, we can also notice that the observed cluster dominance at finite temperature is induced by the entropic gain induced by the cluster centre-of-mass motion, that was not considered in Ref.~\cite{Schuck4}.

\subsubsection{Impact of the rearrangement term} \label{Sec:rearrangement}

\begin{figure}
    \centering
    \includegraphics[scale = 0.45]{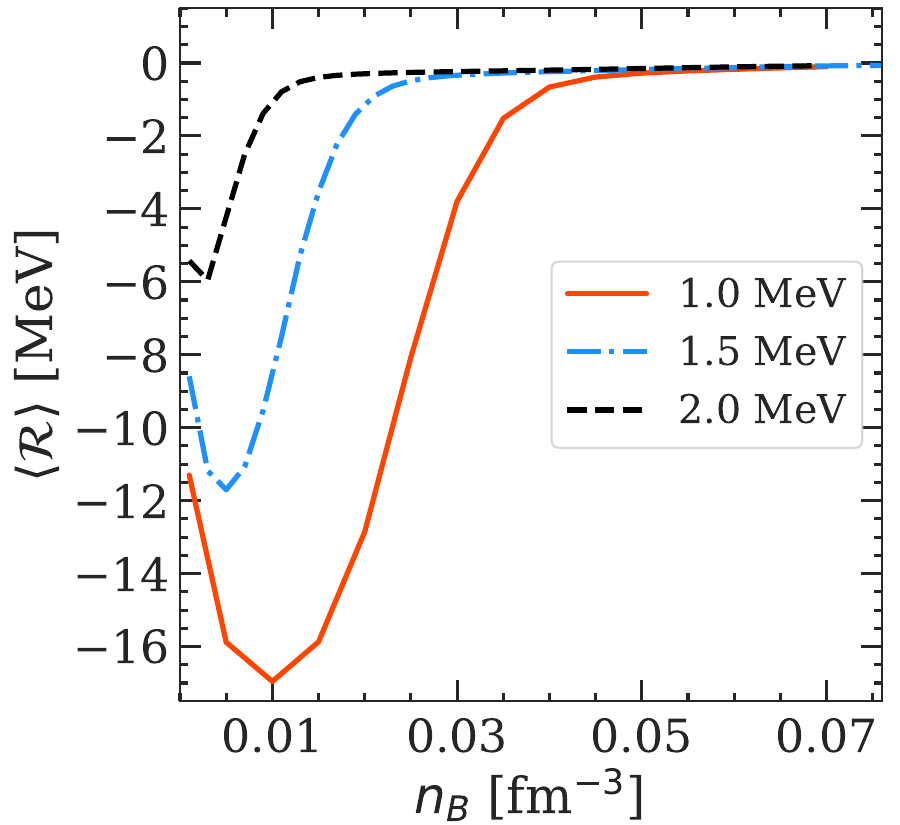}
    \caption{Average value of the rearrangement term $\langle \mathcal{R} \rangle$ as a function of the total baryonic density $n_B$ at three different temperatures: $k_{\rm B}T = 1.0$~MeV (solid red line),  $k_{\rm B}T = 1.5$~MeV (dash-dotted blue line), and  $k_{\rm B}T = 2.0$~MeV (dashed black line). }
    \label{fig:MCP_R_average}
\end{figure}

As discussed in Sect.~\ref{sec:distribution}, due to the dependence of the cluster free energy on the electron gas density via the Coulomb screening term, a rearrangement term, $\mathcal{R}^\j$, appears (see Eq.~(\ref{eq:rearr})). 

To evaluate the importance of the rearrangement term, we performed the MCP calculations neglecting its contribution, Eq.~(\ref{eq:rearr}), in the single-ion canonical potential, Eq.~(\ref{eq:omegaj}).
The resulting nuclear distributions are displayed in Figs.~\ref{fig:MCP_PZ} and \ref{fig:MCP_PA} with orange histograms. 
We can easily see that the impact of the rearrangement term on the MCP distributions is non-negligible, especially at low densities.
Specifically, the distributions obtained without $\mathcal{R}^\j$ (orange) are totally shifted to low $Z$ and $A$, while the correct distributions are peaked for larger nuclei.
However, at high densities, where the nuclear distribution is dominated by light degrees of freedom, the effect from $\mathcal{R}^\j$ appears to be much less important.
This is because the rearrangement term is proportional to the WS volume and the interaction pressure from the Coulomb interaction, see Eq.~(\ref{eq:rearr}). 
When the light clusters are dominant in the crust, the associated WS cell volume as well as Coulomb interaction contribution decrease, thus the rearrangement term in this regions becomes small, not affecting the nuclear distribution. 

This outcome seems to be in contrast with results presented in Ref.~\cite{Grams2018}, where it was shown that the rearrangement term becomes important as density increases (see their Fig.~1).
However, it has to be noted that the conditions explored in Ref.~\cite{Grams2018} are different from those of interest here (more isospin-symmetric supernova matter versus neutron-rich beta-equilibrated PNS crust).
 
To further illustrate this point, in Fig.~\ref{fig:MCP_R_average}, we plot the average value of the rearrangement term, $\langle \mathcal{R} \rangle = \langle V_{\rm WS} \rangle \bar{P}_{\rm int}$, as a function of the baryonic density for three selected temperatures: $k_{\rm B}T = 1.0$~MeV (solid red line),  $k_{\rm B}T = 1.5$~MeV (dash-dotted blue line), and  $k_{\rm B}T = 2.0$~MeV (dashed black line). 
From Fig.~\ref{fig:MCP_R_average}, we can observe that the absolute value of the rearrangement term decreases and approaches zero faster with temperature, which explains the negligible influence of this term in the bottom right panels of Figs.~\ref{fig:MCP_PZ} and \ref{fig:MCP_PA}.

\section{Conclusions}
\label{sec:conlusions}

In this paper, we studied the properties of the PNS inner crust at beta equilibrium in the temperature range from 1~MeV to 2~MeV, where the crust is expected to be in the liquid phase, using both the OCP and MCP approaches.
To this aim, we employed a CLD model approach, in which  the nuclear matter quantities are calculated from finite-temperature mean-field thermodynamics and the surface parameters are optimised consistently with the bulk energy from the fit to the experimental nuclear masses in the AME2016 table \cite{AME2016}.
For illustrative purposes, we have presented the results using the empirical parameters of the BSk24 functional \cite{BSK24}.

In the first part of this study, for each given thermodynamic condition, we calculated the crust composition within the single-nucleus (OCP) approximation using the standard variational minimisation of the total free-energy density. 
Employing four different prescriptions for the ion effective mass, obtained by solving the hydrodynamic equation of the ion moving in a uniform background, we analysed the impact from the translational free energy on the equilibrium configuration. 
We show that when the centre-of-mass motion of the ion is accounted for, the cluster mass and proton numbers are significantly reduced. 
Moreover, by comparing the contribution from the translational term with that from the interface properties, namely Coulomb, surface, and curvature terms, we find that the former becomes more important at higher densities and temperatures, confirming the findings of Ref.~\cite{Dinh2022}. 
As a result, if the NS is sufficiently hot and dependently on the adopted prescription for the ion effective mass, the deeper layers of the PNS crust in the OCP approximation may contain only small clusters, until the crust-core transition is reached.

In a second step, to account for the whole nuclear distributions, we performed fully self-consistent MCP calculations of the PNS crust.
While the solutions in OCP and MCP approaches agree relatively well at low densities, some differences arise, namely, we observe generally a lower (higher) free-neutron (electron) density in the MCP approach with respect to the OCP. 
This discrepancy, which increases with density and temperature, is associated to the non-linear mixing term and the abundance of the light degrees of freedom.  
These latter findings are in agreement with the literature \cite{Souza2009,Hempel2010,gulrad2015,lattimer1991}.

Moreover, we also evaluate the impact on the nuclear distribution of the rearrangement term, which arises from the dependence of the cluster free energy on the electron density through the Coulomb interaction. 
The results show that the rearrangement term affects significantly the distributions of the cluster charge and mass, especially in the conditions where large clusters dominates, highlighting the importance of its inclusion in MCP calculations.

In this work, we only study the co-existence of spherical clusters. 
However, it is important to note that clusters with non-spherical shapes, known as pasta phases, are expected to appear at the bottom of the inner crust. 
Incorporating pasta structures into our MCP approach is left for future investigations.

\begin{acknowledgements}
This work has been partially supported by the IN2P3 Master Project NewMAC, the ANR project `Gravitational waves from hot neutron stars and properties of ultra-dense matter' (GW-HNS, ANR-22-CE31-0001-01), and the CNRS International Research Project (IRP) ``Origine des \'el\'ements lourds dans l’univers: Astres Compacts et Nucl\'eosynth\`ese (ACNu)''.
\end{acknowledgements}


\begin{thebibliography}{}
% etc


\bibitem{hpy2007} P. Haensel, A.~Y. Potekhin, D.~G. Yakovlev, ``Neutron Stars 1. Equation of state and structure'' (Springer, New York, 2007)

\bibitem{Newton2013a} W. G. Newton, K. Murphy, J. Hooker, B.-A. Li, Astrophys. J. Lett. {\bf 779}, 4 (2013)

\bibitem{Horowitz2015} C. J. Horowitz, D. K. Berry, C. M. Briggs, M. E. Caplan, A. Cumming, A. S. Schneider, Phys. Rev. Lett. {\bf 114}, 031102 (2015)

\bibitem{Lin2020}  Z. Lin, M. E. Caplan, C. J. Horowitz, C. Lunardini, Phys. Rev. C {\bf 102}, 045801 (2020)

\bibitem{Schmitt2018} A. Schmitt, P. Shternin, in ``The Physics and Astrophysics of Neutron Stars'', edited by L. Rezzolla, P. Pizzochero, D. I. Jones, N. Rea, and I. Vida\~{n}a, Astrophysics and Space Science Library, Vol. 457, p. 455-574 (Springer, Berlin, 2018)

\bibitem{Chamel2008} N. Chamel, P. Haensel, Living Rev. Relativ. {\bf 11}, 10 (2008)

\bibitem{Goriely2011} S. Goriely, N. Chamel, H.-T. Janka, J. M. Pearson, Astron. Astrophys. {\bf 531}, A78 (2011)


\bibitem{Oertel2017} M. Oertel, M. Hempel, T. Kl\"ahn, S. Typel, Rev. Mod. Phys. {\bf 89}, 015007 (2017)

\bibitem{Fantina2020} A. F. Fantina, S. De Ridder, N. Chamel, F. Gulminelli, Astron. Astrophys. {\bf 633}, A149 (2020)

\bibitem{Carreau2020b} T. Carreau, A. F. Fantina, F. Gulminelli, Astron. Astrophys. {\bf 640}, A77 (2020)

\bibitem{Pons2013} J.~A. Pons, D. Vigan\`o, N. Rea, Nature Physics {\bf 9}, 431 (2013)

\bibitem{Vigano2013} D. Vigan\`o, N. Rea, J.~A. Pons, R. Perna, D.~N. Aguilera, J.~A. Miralles, Mon. Not. Royal Astron. Soc. {\bf 434}, 123 (2013)

\bibitem{Dinh2022_mcp} H. Dinh Thi, A. F. Fantina, F. Gulminelli, submitted to Astron. Astrophys. (2023)

\bibitem{Burrows1984} A. Burrows, J. M. Lattimer, Astrophys. J. {\bf 285}, 294 (1984)

\bibitem{Souza2009} S. R. Souza, A. W. Steiner, W. G. Lynch, R. Donangelo, M. A. Famiano, Astrophys. J. {\bf 707}, 1495 (2009)

\bibitem{Hempel2010} M. Hempel, J. Schaffner-Bielich, Nucl. Phys. A {\bf 837}, 210 (2010)

\bibitem{Blinnikov2011} S.~I. Blinnikov, I.~V. Panov, M.~A. Rudzsky, K. Sumiyoshi, Astron. Astrophys. {\bf 535}, A37 (2011)

\bibitem{gulrad2015} F. Gulminelli, Ad. R. Raduta, Phys. Rev. C {\bf 92}, 055803 (2015)

\bibitem{lattimer1991} J.~M. Lattimer, F.~D. Swesty, Nucl. Phys. A {\bf 535}, 331 (1991)

\bibitem{Shen1998} H. Shen, H. Toki, K. Oyamatsu, K. Sumiyoshi, Nucl. Phys. A {\bf 637}, 435 (1998)

\bibitem{OConnor2007} E. O'Connor, D. Gazit, C. Horowitz, A. Schwenk, N. Barnea, Phys. Rev. C {\bf 75}, 055803 (2007)

\bibitem{Avancini2012} S. S. Avancini, C. C. Barros, L. Brito, S. Chiacchiera, D. P. Menezes, C. Provid\^{e}ncia, Phys. Rev. C {\bf 85}, 035806 (2012)

\bibitem{Avancini2017} S. S. Avancini, M. Ferreira, H. Pais, C. Provid\^{e}ncia, G. R\"{o}pke, Phys. Rev. C {\bf 95}, 045804 (2017)

\bibitem{Pais2015} H. Pais, S. Chiacchiera, C. Provid\^{e}ncia, Phys. Rev. C {\bf 91}, 055801 (2015)

\bibitem{Custodio2021} T. Cust\'odio, H. Pais, C. Provid\^{e}ncia, Phys. Rev. C {\bf 104}, 035801 (2021)

\bibitem{Dinh2022} H. Dinh Thi, A. F. Fantina, F. Gulminelli, Astron. Astrophys. {\bf 672}, A160 (2023)

\bibitem{Carreau2019} T. Carreau, F. Gulminelli, J. Margueron, Eur. Phys. J. A {\bf 55}, 188 (2019)

\bibitem{Carreau2020} T. Carreau, F. Gulminelli, N. Chamel, A.~F. Fantina, J.~M. Pearson, Astron. Astrophys. {\bf 635}, A84 (2020)

\bibitem{dinh2021} H. Dinh Thi, T. Carreau, A. F. Fantina, and F. Gulminelli, Astron. Astrophys. {\bf 654}, A114 (2021)

\bibitem{dinhEPJA21} H. Dinh Thi, A. F. Fantina, F. Gulminelli, Eur. Phys. J. A {\bf 57}, 296 (2021)

\bibitem{Grams2022} G. Grams, J. Margueron, R. Somasundaram, N. Chamel, S. Goriely, J. Phys. Conf. Ser. {\bf 2340}, 012030 (2022)

\bibitem{MagHen2002} P. Magierski, P.-H. Heenen, Phys. Rev. C {\bf 65}, 045804 (2002)

\bibitem{Grill2011} F. Grill, J. Margueron, N. Sandulescu, Phys. Rev. C  {\bf 84}, 065801 (2011)

\bibitem{Baldo2007} M. Baldo, E. E. Saperstein, S. V. Tolokonnikov, Eur. Phys. J. A {\bf 32}, 97 (2007)

\bibitem{Gogelein2007} P. G\"{o}gelein, H. M\"{u}ther, Phys. Rev. C {\bf 76}, 024312 (2007)

\bibitem{Shelley2020} M. Shelley, A. Pastore, Universe {\bf 6}, 206 (2020)

\bibitem{Pearson2018} J. M. Pearson, N. Chamel, A. Y. Potekhin, A. F. Fantina, C. Ducoin, A. K. Dutta, S. Goriely, Mon. Not. R. Astron. Soc. {\bf 481}, 2994 (2018)

\bibitem{Pearson2020}  J. M. Pearson, N. Chamel, A. Y. Potekhin, Phys. Rev. C {\bf 101}, 015802 (2020)

\bibitem{Pearson2022}  J. M. Pearson, N. Chamel, Phys. Rev. C {\bf 105}, 015803 (2022)

\bibitem{Shelley2021}  M. Shelley, A. Pastore, Phys. Rev. C {\bf 103}, 035807 (2021)

\bibitem{Mallik2021} S. Mallik, F. Gulminelli, Phys. Rev. C {\bf 103}, 015803 (2021)

\bibitem{lattimer1985} J. M. Lattimer, C. J. Pethick, D. G. Ravenhall, D. Q. Lamb, Nucl. Phys. A {\bf 432}, 646 (1985)

\bibitem{Ducoin2007} C. Ducoin, Ph. Chomaz, F. Gulminelli, Nucl. Phys. A {\bf 789}, 403 (2007)

\bibitem{Margueron2018a} J. Margueron, R. Hoffmann Casali, F. Gulminelli, Phys. Rev. C {\bf 97}, 025805 (2018)

\bibitem{Margueron2018b} J. Margueron, R. Hoffmann Casali, F. Gulminelli, Phys. Rev. C {\bf 97}, 025806 (2018)

\bibitem{BSK24} S. Goriely, N. Chamel, J.~M. Pearson, Phys.~Rev.~C {\bf 88}, 024308 (2013)

\bibitem{Ravenhall1983_prl} D.~G. Ravenhall, C. J. Pethick, J. R. Wilson, Phys. Rev. Lett. {\bf 50}, 2066 (1983)

\bibitem{Pethick1995} C.~J. Pethick, D.~G. Ravenhall, Annu. Rev. Nucl. Part. Sci. {\bf 45}, 429 (1995)

\bibitem{Maru2005} T. Maruyama, T. Tatsumi, D.~N. Voskresensky, T. Tanigawa, S. Chiba, Phys.~ Rev.~ C {\bf 72}, 015802  (2005)

\bibitem{Newton2013} W.~G. Newton, M. Gearheart, B.-A. Li, Astrophys. J. Suppl. Series {\bf 204}, 9 (2013)

\bibitem{Ravenhall1983} D.~G. Ravenhall, C.~H. Pethick, J.~M. Lattimer, Nucl. Phys. A {\bf 407}, 571 (1983)

\bibitem{AME2016} M. Wang, G. Audi, F.~G. Kondev, W.~J. Huang, S. Naimi, X. Xu, Chinese Phys. C {\bf 41}, 030003 (2017) and Atomic Mass Data Center,  http://amdc.in2p3.fr/web/masseval.html

\bibitem{Avancini2009} S. S. Avancini, L. Brito, J. R. Marinelli, D. P. Menezes, M. M. W. de Moraes, C. Provid\^{e}ncia, A. M. Santos, Phys. Rev. C {\bf 79}, 035804 (2009)

\bibitem{Shen2011} H. Shen, H. Toki, K. Oyamatsu, K. Sumiyoshi, Astrophys. J. Suppl. Ser. {\bf 197}, 20 (2011)

\bibitem{Schneider} A. S. Schneider, L. F. Roberts, C. D. Ott, Phys. Rev. C {\bf 96}, 065802 (2017)

\bibitem{Lattimer1985} J. M. Lattimer, C. J. Pethick, D. G. Ravenhall, D. Q. Lamb, Nucl. Phys. A {\bf 432}, 646 (1985)

\bibitem{Epstein1988} R.~I. Epstein, Astrophys. J. {\bf 333}, 880 (1988)

\bibitem{Sedrakian1996} A. D. Sedrakian, Astrophys. Space Sci. {\bf 236}, 267 (1996)

\bibitem{Magierski2004} P. Magierski, A. Bulgac, Nucl. Phys. A {\bf 738}, 143 (2004)

\bibitem{Magierski2004b} P. Magierski, A. Bulgac, Acta Phys. Polon. B {\bf 35}, 1203 (2004)

\bibitem{Martin2016} N. Martin, M. Urban, Phys. Rev. C {\bf 94}, 065801 (2016)

\bibitem{Chamel2017} N. Chamel, J. Low Temp. Phys. {\bf 189}, 328 (2017)

\bibitem{BBP1971}  G. Baym, H. A. Bethe, C. J. Pethick, Nucl. Phys. A {\bf 175}, 225 (1971)

\bibitem{Grams2018} G. Grams, S. Giraud, A. F. Fantina, F. Gulminelli, Phys. Rev. C {\bf 97}, 035807 (2018)

\bibitem{Barros2020}  C. C. Barros, D. P. Menezes, F. Gulminelli, Phys. Rev. C {\bf 101}, 035211 (2020)

\bibitem{Pelicer2021} M. R. Pelicer, D. P. Menezes, C. C. Barros, F. Gulminelli, Phys. Rev. C {\bf 104}, L022801 (2021)

\bibitem{Botvina2010} A. S. Botvina, I. N. Mishustin, Nucl. Phys. A {\bf 843}, 98 (2010)

\bibitem{Schuck3} 
B. Zhou, Y. Funaki, H. Horiuchi, Z. Ren, G. R\"{o}pke, P. Schuck, A. Tohsaki, C. Xu, T. Yamada, Phys. Rev. Lett. {\bf 110}, 262501 (2013)

\bibitem{Schuck1} P. Schuck, T. Sogo, G. R\"{o}pke, Prog. Th. Phys. Suppl. {\bf 196}, 56 (2012)

\bibitem{Schuck2} G. C. Strinati, P. Pieri, G. R\"{o}pke, P. Schuck, M. Urban, Phys. Rep. {\bf 738}, 1 (2018)

\bibitem{Schuck4} T. Sogo, G. R\"{o}pke, P. Schuck, Phys. Rev. C {\bf 82}, 034322 (2010)








\end{thebibliography}


\end{document}